\documentclass[12pt]{article}
\usepackage{epsfig}
\begin{document}
           
\title{Stochastic Lattice Models for the Dynamics of Linear Polymers}
\author{J.M.J. van Leeuwen \\
Instituut-Lorentz, University of Leiden, P.O.Box 9506, \\
2300 RA Leiden, the Netherlands\\*[4mm]
Andrzej Drzewi\'nski \\
Institute of Physics, University of Zielona G\'ora, \\
Prof. Z. Szafrana 4a, 65-516 Zielona G\'ora, Poland}
\maketitle

\begin{abstract}
Linear polymers are represented as chains of hopping reptons and their motion 
is described as a stochastic process on a lattice. This admittedly 
crude approximation still catches essential physics of polymer motion, i.e. 
the universal properties as function of polymer length. More than the static properties, 
the dynamics depends on the rules of motion. Small changes in the
hopping probabilities can result in different universal behavior. In particular the
cross-over between Rouse dynamics and reptation is controlled by the types 
and strength of the hoppings that are allowed.

The properties are analyzed using a calculational scheme based on an analogy with 
one-dimensional spin systems. It leads to accurate data for intermediately long polymers.
These  are extrapolated to arbitrarily long polymers, by means of finite-size-scaling 
analysis. Exponents and cross-over functions for the renewal time and the diffusion 
coefficient are discussed for various types of motion.
\end{abstract}
\setcounter{section}{-1}
\section{Introduction}\label{intro}

Long polymers are strongly interacting many-body systems. They are partly stiff and partly
flexible and have a very large number of degrees of freedom. Not only the interaction between 
the monomers is important, but also the interaction with the medium in which they are immersed, 
which makes them notoriously difficult systems for a quantitative theoretical analysis from first 
principles. If the properties of polymers were not so interesting and important for 
applications and if polymer motion was not so vital for processes in living material, 
their study would never have been taken up by theoretical physicists.
The eternal dilemma is to choose between being realistic and keeping it simple. While the 
culture of chemists opts for being realistic and deals with specific properties, 
the culture of physicists tends towards simplicity and aims at generic properties. 
This review is in the tradition of physics and surveys what one can learn from simple 
models for the generic dynamics of polymers.

As a classical system, polymer behavior can well be studied by computer simulation.
Polymers physics has highly benefited from computer simulations and with the 
continuing growth of computer power and memory size it will keep gaining insight from 
simulations in the future \cite{Binder}. 
Nevertheless, the size of the polymers and the intricacy of the molecular motion severely 
limits simulations to relatively short polymers and short simulation times, while the 
most physically interesting properties are for long polymers and long times. 
Thus alternatives in the numerical study of polymers, which complement and articulate 
the simulation results, are most welcome. In the past decades the study of polymer motion 
has become a successful enterprise, due to a number of simplifications of which we mention 
a few that are important for this review.\\

{\it \large Scaling and criticality}  

The first observation, which brought theoretical physicists into polymer research,
was made by de Gennes \cite{deGennes1}. He demonstrated in his book 
{\it Scaling Concepts in  Polymers Physics} 
the possibility of a systematic study of the properties of polymers as function 
of the number $N$ of constituing monomers. Such properties do not depend on details 
of the basic units and their interaction. The dependence on $N$ is powerlike, 
which is the signature of critical behavior. Indeed de Gennes could show that a 
long polymer is in a critical state. The critical point is the limit $N \rightarrow \infty$.
Whereas most systems have to be carefully tuned to become critical, polymers just 
have to be long in order to demonstrate critical behavior. This is a profound observation.
As a consequence, all the powerful field-theoretical methods, developed for critical 
phenomena, can and have been applied to polymers. Consider as  illustration  the amazing 
result, that the properties of self-avoiding polymers, can be found from a field theory 
where the number of components vanishes. Truely a triumph of the abstract methods of 
theoretical physics! \cite{deGennes1} Before renormalization theory came into full swing, 
such a result was already anticipated through a map of polymer configurations on random 
walks on a lattice. For long random walks the lattice structure becomes irrelevant and 
the study of random walks on a lattice showed that the end-to-end distance only depends 
on such global aspects as the dimension of the lattice and on whether the walk is 
self-avoiding or not \cite{Pelissetto}.\\

{\it \large Lattice Dynamics}

The second point is that polymer behavior can be elucidated by lattice models.
The use of lattice models in the description of polymers has a long tradition. Already
in 1939 Meyer \cite{Meyer} used a lattice to represent polymer configurations. Verdier 
and Stockmayer \cite{Verdier} performed in 1962 the above mentioned self-avoiding 
random walks on a lattice. For a comprehensive review of polymer
chains as walks on a lattice see \cite{Vanderzande}.
In 1981 Edwards and Evans \cite{Edwards} proposed the 
cage model for the reptative motion of polymers dissolved in a gel. 
The use of lattice models for dynamical properties obtained a further boost 
by the work of Rubinstein \cite{Rubinstein}, who introduced a very simple lattice model 
for reptation.  An important extension was given by Duke \cite{Duke1,Duke2,Duke3}, 
who incorporated an electric driving-field as a bias in the hopping rules. 
This allows to describe {\it gel-electrophoresis}, the basic technique in DNA 
fingerprinting \cite{Zwolak}. The combination of the Rubinstein and Duke hopping 
rules leads to the Rubinstein-Duke (RD) model. It is the paradigm of the reptative motion 
of polymers dissolved in a gel. The RD model has been lucidly analyzed by Widom et al.
\cite{Widom}, who gave also informative illustrations for short polymers. In this paper 
we will frequently return to the RD model for reference.

Mapping a polymer on a lattice, implies that the motion of the polymer has to be
described as a stochastic process. The incompatibility of a lattice description with
continuous motion is not the only reason to introduce a stochastic element. If one 
eliminates the influence of the environment in order to get a single polymer problem,
the motion of the polymer will automatically be stochastic.\\

{\it \large Analogy with the quantum mechanics}

A third stimulating input in the polymer field comes from experience with quantum 
problems. Stochastic processes are governed by the Master Equation, which has a 
mathematical structure similar to the Schr\"odinger equation. One can
exploit the analogy between the two equations, although they
refer to totally different physical situations.
The Master Equation describes the temporal evolution of the probability distribution, 
while the Schr\"odinger equation gives that of the wavefunction.  The main difference is 
that in the Schr\"odinger equation, time is combined with the imaginary unit $i$ , 
leading to generally complex wavefunctions of which the absolute square
has a probability interpretation. Another important difference is that
the Schr\"odinger operator is hermitian, implying a spectrum of real eigenvalues. The
Master operator is real but in general not symmetric (or non-hermitian), 
such that the eigenvalues are not necessarily real. From the physics behind the 
Master Equation one knows however, that all the eigenvalues 
have a negative real part, since all the solutions of the Master Equation have to 
decay towards the asymptotic stationary solution. In this analogy the
asymptotic solution corresponds to the groundstate of the Schr\"odinger equation.
Conservation of probability implies that the Master operator has a zero eigenvalue and
the corresponding (right) eigenfunction gives the probability distribution of the 
stationary state. In spite of these different interpretations, the solution technique for 
the Schr\"odinger equation can be successfully employed in solving the Master Equation.\\

{\it \large How to use the simplifications}
 
Let us now comment in more detail on how one can take advantage of these simplifications 
for the analysis of polymer motion.

{\it Criticality}. The attractive aspect of critical systems is that the properties are universal, i.e.
independent of the microscopic details of the ingredients. This allows one to succesfully
study intricate polymers by simple models having the essential features of
the motion. As far as the limit  $N \rightarrow \infty$ is concerned, only the 
field-theoretical analysis fully benefits from this limit. 
All other methods tend to analyze systems with $N$ as large as possible, 
hoping to see the asymptotic behavior. A powerful instrument in this extrapolation is 
finite-size-scaling analysis, a well-developed theory in critical phenomena. 
One fits the finite-$N$ data with formulae anticipating the 
asymptotic behavior. These finite-size corrections considerably sharpen up the results,
but require very accurate data for finite systems and those obtained by 
simulation often have too large statistical errors to be useful.
Another clarifying notion, due to critical theory, is cross-over scaling. It handles 
situations where competing tendencies in asymptotic behavior are present, 
by designing combinations of variables in which a unifying picture can be formulated. 

{\it Lattices}. Polymer motion can hardly be described as a stochastic process on a lattice on monomer
level. The structure of a polymer on monomer level is 
partially stiff and partially flexible. The distance between the monomers is fixed, 
but the links between the monomers have a rotational degree of freedom, only the angle 
between two successive links is stiff. This means that nearby monomers are highly
correlated, but the correlation between links is lost after a number of steps, 
the {\it persistence length}, depending on the stiffness of the chain. It was  observed that 
the motion of polymers could successfully be discussed by the introduction \cite{Doi} of 
the notion of {\it reptons}. A repton, or Kuhn's unit, is a group of monomers of the order 
of the persistence length. It can include up to 50 monomers. Two successive reptons 
along the chain move more or less independently with one important restriction: their
mutual distance may not exceed a certain maximum length, otherwise the polymer
would be stretched beyond its possibilities. Taking the repton as basic unit of motion,
the polymer motion can be reduced to hopping reptons on a lattice. There is 
still a large variety of possible hopping rules, which go under names such as ``bond
fluctuation model'', ``cage model'' and ``repton model''. 
 \begin{itemize}
\item {\it Bond-fluctuation models}. The basic unit represents rather a monomer than
a repton. It  occupies a set of lattice points, e.g. the 8 corners of a cube of a simple 
cubic lattice.  The units fully exclude each other and the links between the units are 
restricted to a set of distances. This model contains the self-avoiding character of the 
monomers. The mutual exclusion implies interactions between distant parts of the chain 
and practically only simulations can reveal the properties of the model. For simulations
with the bond fluctuation method see \cite{Kremer,Binder}.
\item {\it Cage models}. They are designed for describing reptation as the hopping of a 
defect (later to be called a hernia) along the chain. Here the units are reptons, which are 
loose structures with less tendency to exclude each other. Most information on these models 
has again been obtained by simulation, but we will show that also analytical 
methods can be used to study their properties
\item {\it Repton models}. The RD model is the prototype of this class. The units are reptons
and the main type of motion is the diffusion of stored length along the chain. 
The properties of repton models are quite similar to those for cage models. They differ in the 
representation of the stored length, which is in the cage model a ``hernia'' and in the 
repton model a slack link.
\end{itemize}
Besides these main lattice models, other models are proposed, e.g. the recently introduced 
``necklace''model, which views the chain as a succession of beads and holes \cite{Terranova}.
In this review we focus on the cage model and the RD model, because they can easily be
described by a common set of hopping rules, which facilitates the comparison of the results
for the two models.

Discretizing space is a common practice and the standard 
justification relies on the fact that a continuum description follows from a finer and 
finer grained lattice. This may be true, but it is not the spirit of our approach. We 
take lessons from the lattice gas version of a fluid. The most successful (and simplest) 
lattice gas is the one in which there is a hardcore exclusion in the cell and only an 
interaction between nearest-neighbors cells (the Ising version). 
More refined lattices introduce a mixing of packing problems and 
particle motion, which is hard to disentangle. Similarly we tune the lattice constant 
such that  successive reptons are either in the same cell or in neighboring cells. 
By excluding further distances, the size of the lattice constant has to be of the order 
of the reptons themselves, i.e.~of the persistence length. 

This review is confined to the use of lattice models in understanding polymer
behavior, which is admittedly a severe restriction. It excludes e.g. the study of specific 
polymers, which are determined by their specific chemical composition. Using  
coarse-graining notions as  ``reptons'' and ``hopping rules'',  washes out specific details. 
Nevertheless, as argued above, generic properties, such as the large $N$ behavior,  
can be fruitfully investigated by lattice models.

{\it Quantum analogy}. The analogy between the Master Equation and the Schr\"odinger equation 
has been noticed for quite a while, but mapping one difficult problem onto another difficult 
problem does not bring the solution any closer.  The Master Equation for polymer motion 
has, however, an aspect which makes the analogy
with quantum systems very attractive: its linear structure maps it to a one-dimensional
quantum system with only nearest neighbor interactions. This follows from the fact
that the hopping of reptons is only hindered by the position of its neighboring reptons (as
long as self-avoidance is neglected). In the past decade a powerful numerical technique 
has been developed by White \cite{White1,White2} for one-dimensional quantum systems. 
Whereas in quantum mechanics the restriction to one-dimensional systems is artificial 
(and the extension to higher dimensional systems tedious and less successful), the linear 
structure is a natural aspect of the polymer problem. The application of White's
method to the polymer chain is the core of this review.\\

{\it \large Aim of this review}

The literature on polymer motion is enormous and several excellent review papers exist 
\cite{Slater,Rubinstein1,vanHeukelum}, notably the review of Viovy \cite{Viovy2}, 
which focusses on gel-electrophoresis, reviewing
many experimental results and theoretical notions. It contains a section 
``Investigating the repton model in depth'' and the present review may be seen as an 
elaboration of that section. So we are concerned with the theoretical aspects of lattice 
models describing the polymer motion as a stochastic process of hopping reptons. 
In this review universality is the key issue, i.e.~the dependence of the properties
on the length $N$ of the polymer chain. On the one hand universality gives us the liberty to 
study simplified models in the idea that the universal properties are largely model 
independent. This motivation is similar to the study of critical phenomena where 
indeed the universality classes are very large and only such aspects as the dimensionality
of the lattice and the symmetry of the order parameter influence the universality class. 
On the other hand we will find that small changes in the hopping rules of the reptons can 
change the universal properies of polymer dynamics. The dimension of the embedding 
lattice is less important for the universal properties than in the case of critical phenomena. 
The one-dimensional character of the chain makes the motion rather 
independent of the embedding lattice. 

The number of properties of polymers which can be studied is also bewildering.
The calculational method that we use, limits us to those close to the stationary state. 
Two coefficients, which are intimately related, stand out: the renewal time $\tau$ and the 
(zero field) diffusion constant $D$. The first measures the time needed for a transition 
to a new configuration. So it concerns the decay of the slowest mode towards the 
stationary state.
The second is a measure for the mean square displacement in the stationary state. 
Both quantities show power-law behavior in the limit of long polymers and 
they vary with the type of hoppings that are included. 
It is this variation that can be demonstrated clearly using lattice dynamics.

The calculational method that we use practically forbids to include self-avoidance.
Self-avoiding chains feel a long-range interaction between the reptons, which 
is at odds with the treatment of short-range quantum chains.
Self-avoidance is a conceptually very important element for universality and 
it makes the form of criticality really intriguing with exponents that are not rational.  
But on repton level the mutual exclusion is less severe than on monomer level
 since the reptons are loose structures. 
Also there are situations where the polymer behaves as an 
ideal chain such as in polymer melts \cite{deGennes1}. Thus the polymers
that we consider here are ideal and flexible chains of reptons. \\

{\it \large Layout of the paper}

In this review we discuss lattice models for which the dynamics is governed
by the rules to be defined below. To make the review more easily accessible we sketch 
the aim and content of the chapters to follow. 
\begin{enumerate}
\item We start out to briefly describe the {\bf context} of these lattice models by stressing
the influence of the environment of the polymer chain. This gives the role of  the
various hopping rates that are defined in the next chapter and what may and may not be
expected from a lattice description of motion.
\item In the chapter dealing with the {\bf model} we define the chain configurations,
the hopping rates and their role. The driving field is introduced and the two main models: 
the repton model and the cage model are described.
\item The {\bf Master Equation} forms the basis for all the calculations to follow.
The important aspects of this equation, e.g. the stationary state, gap and detailed balance, 
are discussed in some detail. Special attention gets the idea of contracting the Master
Equation to a more coarse grained description. This is an important notion and it pays
off to have the procedure explicitely formulated, since it is used several times later on.
\item The chapter on {\bf correlations} deals with the definitions of the correlations in the
stationary state, both for the link probabilities and for the repton velocities. It turns out
that three velocities are needed: the drift velocity, the curvilinear velocity and the 
electrical current. 
\item Since we calculate the diffusion coefficient as the linear response to a field we
discuss in a separate chapter the {\bf linearization} of the Master Equation. It gives
the surprising result that the full probability distribution for any charge distribution,
can be related to that of a special charge distribution: the magnetophoresis case, 
where only the head repton of the chain is charged. This is a powerful instrument 
to interrelate the correlations of the various types of chains.
\item Few {\bf exact results} can be derived
directly from the Master Equation. A number of models become exactly soluble
in a one-dimensional embedding lattice. Although these exactly
soluble models do not impressively contribute to the understanding of polymer dynamics,
we will pay a fair amount of attention to them, because they serve as checks on ideas
and on the accuracy of calculational schemes for more complicated situations.
\item As mentioned earlier the core of our calculations
is based on the Density Matrix Renormalization Group ({\bf DMRG}) method for establishing 
the probability distribution  in the stationary  state of the chain. As background
information, we outline in this chapter the strategy of the procedure in relation
to our reptating chains. For a more elaborate description of the method we refer to
the many reviews on this method.
\item One of the virtues of the DMRG results is that they allow a {\bf finite-size analysis}
for both the renewal time $\tau$ and the diffusion coefficient $D$. We show
typical results for the RD model.
\item The special features of our extensions of the RD model come to life, when
considering {\bf cross-over} from reptation to Rouse dynamics. It is a general feature
of the inclusion of constraint release in the hopping rules. 
\item More detailed information than contained in $\tau$ and $D$ follows from considering
{\bf local orientation}, for weak and stronger fields. The electrophoresis case has 
an informative  structure for weak fields, whereas the
magnetophoresis case has a delicate shape for stronger fields. 
\item The last chapter is devoted to {\bf remaining problems}. It deals with shortcomings
of the models discussed here, problems that yet could be analyzed in the present setting
and the intriguing issue of the transition from random to oriented chains.  
\end{enumerate}

\section{The Physical Context} \label{physics}

The diffusion coefficient $D$ and the renewal time $\tau$ have a power-law
dependence on the number of reptons $N$ \cite{deGennes2}. On physical grounds one expects
that these power laws are interrelated as
\begin{equation} \label{0}
D \tau \sim R^2_g.
\end{equation}
Here $R_g$ is the radius of gyration or in more operational terms the 
average end-to-end distance of the chain. The rational behind this relation
is that the chain renews itself when it drifts over a distance $R_g$.
Since this drift is diffusive, it takes the time $R^2_g/D$. Depending on 
the circumstances each of the three entries in (\ref{0}) is governed 
by an exponent, which gives the power in the dependence on $N$. That 
for the radius of gyration is standardly denoted by $\nu$. Considering 
the polymer as a random walk in space one finds $\nu=1/2$. Due to 
self-avoidance the exponent changes to $\nu=0.5877 \pm 0.0006$ 
(best theoretical estimate for $d=3$ \cite{Madras};
for $d=2$ the exact value is $\nu = 3/4$). One would think that always some 
self-avoidance is present, also on repton level, but the situation is
more complicated as we will indicate. The renewal time is associated
with the viscosity and the diffusion coefficient can be measured in
the standard way. Note that all three quantities refer to a single polymer.

The main issue in polymer physics is whether the properties of the system
can be deduced from the behavior of a single polymer in a well chosen 
surrounding. This depends of course on the environment of the polymer.
We briefly discuss here the typical circumstances for dilute solutions, gels
and melts and their relation to lattice dynamics.\\                                                                                                              

{\it \large Dilute solutions}

On one side of the spectrum is  a 
polymer chain dissolved in a good solvent. Here the chain units 
(reptons) are free to move in all directions, only constrained
by the integrity of the chain, which means that the reptons
may not separate too far apart, otherwise the chain would break.
Standarly the constraining forces are represented by a harmonic potential.
In a good solvent the polymer drags the solvent along which in turn
leads to an hydrodynamical interaction between the parts of the 
polymer. The reptons move collectively rather than independently. 
This is described by the Zimm model \cite{Zimm} which predicts
amongst others that $D \sim N^{-\nu}$. This  is sometimes taken as
a way to measure $\nu$ \cite{Smith}. 

If the solvent is highly viscous
the dynamics is well described by the Rouse model for the dynamics 
\cite{Rouse}. Here the mutual interaction between parts of the chain
via the environment can be ignored.
The Rouse model has the advantage that all the modes of the
reptons can be explicitly calculated. One finds $D \sim N^{-1}$ and 
$\tau \sim N^2$ and consistently with (\ref{0}) that $\nu=1/2$
(since self-avoidance is absent). In a poor solvent the statistics 
of a single polymer is dominated by the background and not treatable 
by lattice dynamics. From the viewpoint of lattice dynamics only the
Rouse dynamics can be described by lattice models since the 
environment of the polymer is inert in the Rouse description. \\ 

{\it \large Gels}

On the other side of the spectrum is a polymer dissolved in a gel. A gel
is an open rigid structure with pores through which the polymer may find its way. 
The gel provides a tube in which the chain may move, like a snake, 
only longitudinally along the contour traced out by the tube. 
This form of motion is known as {\it reptation}. The steps of reptation consist of
an accumulation of stored length followed by diffusion of the length along the tube.
The polymer can only change its confining tube at the ends of the chain.  
New elements of the tube are created when
end points of the chain find new pores and similarly parts of the tube are
annihilated when the segments retract inside the tube. Therefore 
the tube configuration is slowly varying, by motions of the
ends of the chain, which are  called {\it contour-length fluctuations} (CLF). 
The interplay between the internal diffusion of stored length and the external 
contour-length fluctuations is the essence of the reptation. 
The chain cannot extend the tube if not sufficient
stored length is accumulated near the end, nor will it retract inside the tube if 
too much stored length is present there. Without contour-length fluctuations the
tube would be an invariant and the chain would not drift in space.
One of the virtues of the RD model is that it incorporates this interplay.
It is therefore a simple model describing reptation. 

Applying an electric field as driving force yields an overall drift in the direction 
of the field. This is called {\it gel-electrophoresis}, an important ingredient in 
such techniques as DNA sequencing \cite{Dolnik}. Assume for the moment that
the chain is oriented from tail to head in the direction of the field. Then the
bias by the field tends the chain to retract at the tail side. This accumulation
of stored length at the tail  side diffuses and is driven along the tube to the head 
where it may create an extension of the tube. As a consequence the chain as a whole
has moved one step in the direction of the field.
No long calculation is needed to convince oneself that this is a slow
process, the slower the longer the chain. It is however far from trivial to
figure out the precise length dependence of the resulting drift velocity.
De Gennes predicted for reptation that $D \sim N^{-2}$ and $\tau \sim N^3$ \cite{deGennes1}.  
One already sees a complication from the fact that both ends are equivalent.
Not necessarily one end of the tube will dominantly shrink and the
other grow. This would be the case when the chain is permanently ordered
one way in the field. Certainly for very weak fields, the orientation of 
the chain will vary in time: head and tail interchange their role.
Increasing the field strength or the length of the chain, this tumbling over 
becomes very rare and the probability distribution develops two peaks 
corresponding to the two orientations. 
For even stronger fields the chain diffusion and field force may work against 
each other. This happens when the chain gets hooked around
an obstacle. The stored length has to drift for a long section against the field 
and the drift will drop because the barrier for fluctuations to
disentangle the chain from the obstacle becomes too large.
Thus an intricate competition between field
strength and chain length may be expected.\\

{\it \large Melts}

In between these two extremes of the spectrum are polymer melts. 
In the two previous cases one 
could restrict oneself to the motion of a single polymer, provided the solution is
sufficiently dilute. In a melt one has a multi-polymer problem since the
polymers constantly interact with each other. As a sort of mean field approximation, 
one can take the interaction with the surrounding polymers into account
by representing them as a confining tube for the polymer in consideration.
This idea due to Doi and Edwards \cite{Doi} reduces the multi-polymer problem again
to a single-polymer problem.  The properties of the confining tube have recently been 
quantified by Zhou and Larson \cite{Zhou}, who calculated, by molecular dynamics,
the confining potential. They find that the tube diameter varies
with the time scale on which the dynamics is observed. This shows that the tube 
created by the others is not a rigid structure, like a gel. So one has to include 
transverse motion in the tube as a form of {\it constraint release} (CR). 

The melt is the most important and challenging case. The general observation is
that a melt of short polymers exhibits Rouse behavior while the long polymers are
forced to reptate. This phenomenon has been further analyzed by Smith et al. 
\cite{Smith2} and recently by Zamponi et al. \cite{Zamponi}. 
They study the dynamics of a test polymer of length $N$ in a melt 
of polymers of varying length $N_m$. If $N_m \ll N$ the environment will act as
a watery solution. In the other extreme $N_m \gg N$ the environment behaves as
a gel. In between the behavior is dominated by the magnitude of the constraint 
release, which depends on $N_m$. Varying $N_m$ yields cross-over from Rouse dynamics
to reptation. Smith et al. \cite{Smith} study the self-diffusion coefficient of the test
polymer. The signature of cross-over of Zamponi et al. \cite{Zamponi}
is based on the behavior of the dynamic structure function as measured by 
Neutron Spin Echo. The dynamic structure function is sensitive to short time 
correlations. (In contrast, our signature probes the long time behavior, since it is 
based on the diffusion coefficient $D$ and the renewal time $\tau$, see also \cite{Kariyo}).

It is argued that in the melt no effects of self-avoidance are observed, since the polymer 
cannot distinguish the reptons of its own chain from those of the others \cite{deGennes1}. \\

From this sketch it is clear that {\it reptation} is the key notion. To phrase it in
the words of Lodge \cite{Lodge} `` The reptation model is the cornerstone for 
our current understanding of the dynamics of entangled polymers''.
In the cited paper Lodge elucidates some of the persistent discrepancies between
theory and experiment of the asymptotic behavior $\tau$ and $D$.
The problem is on the one hand that the measured
$\tau$ and $D$ do not seem to obey the general relation (\ref{0}). For the melt we 
must take for $D$ the self-diffusion. The measurements of $\tau$ and $D$ are 
reconciled by Lodge \cite{Lodge} through 
new measurements and a re-analysis of the older experiments. 
On the other hand there is also a discrepancy between the value of the theoretical 
renewal exponent for the reptation model and the value from measuring the viscosity
which give $\tau \sim N^{3.4}$ for several decades in $N$. 
A host of suggestions have been put forward to remove or alleviate the 
discrepancy between theory and experiment. 
One class of explanations \cite{Doi1,Rubinstein,Milner,Carlon2,Paessens} suggest that 
the experiments are not observing the ultimate asymptotic behavior and that a 
cross-over to the theoretical values happens outside the experimental regime. 
The other schools blame the reptation model for missing essential elements 
determining the true asymptotic behavior, as observed by experiment. 
E.g. a possible explanation, put forward by Barkema and Panja \cite{Panja}, points
to a failure of the standard treatments to account properly for the interaction
with the neighboring polymers in a melt. This suggests a noticeable change in
the exponent thereby closing the gap between theory and experiment. \\
 
Lattice models are most suited to describe the polymer motion
in  a gel. It is not such a big step to replace the gel, with randomly distributed
pores, by a regular lattice structure. In a lattice the chain traces out a contour
which is randomized by the contour-length fluctuations at the ends of the chain.
After a sufficiently long time, the renewal time, the chain has found a new contour
(tube) which has no memory of the contour from which it started. The RD model
incorporates these contour-length fluctuations, as the hopping end-reptons are
free to embark on new pores (provided stored length allows them), or to leave occupied
pores and move further inside (provided stored length is not prohibiting them).  

Constraint release allows the reptons to move sideways with respect to the confining tube.
These types of hoppings are not included in the RD-model. In this review we pay ample
attention to the effect of constraint release, as competing with contour-length
fluctuations. The interesting case
is when they are small compared to the reptative moves. Then the tube
remains a slow variable, but the renewal time is affected by the contraint 
release. The competion between constraint release and the contour-length 
fluctuations gives cross-over. When contour-length fluctuations are the main
mechanism for renewal, the chain is reptating with the typical reptation 
dependence on the length. When constraint release becomes the faster mechanism
for renewal, the chain starts to display Rouse dynamics with another dependence
of the renewal time on the chain length. The longer the chain, the slower the 
contour-length fluctuation mechanism becomes. So which mechanism is the fastest 
depends on a combination of the constraint release rate and the length ot the chain.
Rouse dynamics will always result for extremely long polymers in a gel, no matter
how small a fixed rate of constraint release. This may seem counter-intuitive and indeed it is
opposite to the observed tendency in melts. The hidden assumption in this cross-over
picture is that the constraint release rate is taken independent of the length of the polymer. 
In a  melt of equal-sized polymers, the constraint release also slows down 
with the length of the polymers and thus the competition can turn in the opposite
direction in the limit of large $N$. 
There are various modes of constraint release 
and even a competition between the types is possible. This makes cross-over
from reptation to Rouse dynamics a very rich phenomenon. 

In this review we mainly concentrate on the renewal time $\tau$ and the diffusion 
coefficient $D$. We calculate them independently such that we are able to
check relation (\ref{0}). The dependence of these quantities on the chain length is a 
signature for reptation or Rouse dynamics, since these forms of motion each have a 
typical power dependence on $N$ for $\tau$ and $D$. Not only the true asymptotic
power is important. For a proper understanding it is equally important to indicate
how large $N$ has to be in order that the true asymptotic exponent emerges. 
The lattice models that we consider here, can give a clear hint where this happens. 

\section{The Lattice Description}\label{lattice}

The reptons are located in cells of a lattice. Many lattices are possible and some have
special properties which simplify or denature the motion. We consider here the
class of $d$-dimensional (hyper)cubic lattices which combine simplicity with regularity. 
In particular the influence of the dimensionality can be analyzed without ambiguities.
Since the self-avoidance is neglected here, the structure of the lattice demonstrates
itself only indirectly. The linear structure of the polymer makes the number $z$ of nearest 
neighbors the main aspect of the embedding dimension. Seeing the chain as a walk 
on the lattice, each step has $z$ possibilities. For (hyper)cubic lattices $z=2d$.

The reptons get a discrete cell coordinate, regardless
of their position in the cell. They are labeled with an index $i$ running from $i=0$,
the {\it tail} repton, to $N$, the {\it head} repton. In total we have $N+1$ reptons. 
The reptons need not be the same along the chain. They can have e.g. different charges. 
If all the reptons are identical, we have a symmetric chain, 
where the choice of head and tail is arbitrary.
A chain configuration is characterised by $N+1$ position variables 
${\bf x}_0, \cdots , {\bf x}_N$, for which one can take the centers of the occupied 
cells. For an allowable sequence, two successive positions 
${\bf x}_i$ and ${\bf x}_{i+1}$ must be either the same 
or a nearest neighbor distance apart. We call the difference 
\begin{equation} \label{b1}
{\bf y}_i = {\bf x}_i - {\bf x}_{i-1},
\end{equation} 
a link. If the two successive reptons $i-1$ and $i$ are in the same cell, the link is 
zero or {\it slack}. Otherwise it is a nearest-neighbor distance and the link is called
{\it taut}. The slack links represent elements of stored length, which are the basic 
ingredients for reptation. But we allow the motion to be more general than reptation. 
The only restriction on the chain is that the string of reptons traces out a {\it tube}, 
which is a sequence of nearest-neighboring cells, generated by the taut links. If a link 
were to connect two cells further apart than a nearest-neighbor distance, it would mean a 
rupture in the polymer. The links determine the configuration of the chain,
since the absolute position of the chain in the lattice is hardly of importance. So the
configurations of the chain are given by the vector 
${\bf Y}=({\bf y}_1,\cdots ,{\bf y}_N)$,
showing that the configuration space is finite and of a one-dimensional structure. 
Each link has the choice of being slack 
${\bf y} =0$ or one of the $2d$ possible taut links. For the taut links we use $d$ unit 
vectors ${\bf e}_\alpha$ in the direction of the axis of the lattice and $d$ vectors 
$-{\bf e}_\alpha$ pointing in the opposite direction. If there is no reason to distinguish 
the two types of vectors, we denote them as ${\bf e}_k$.
The possible number of 
configurations is $(2d+1)^N$, which is a staggering number for long polymers.
\begin{figure}[h]
\begin{center}
    \epsfxsize=12cm
    \epsffile{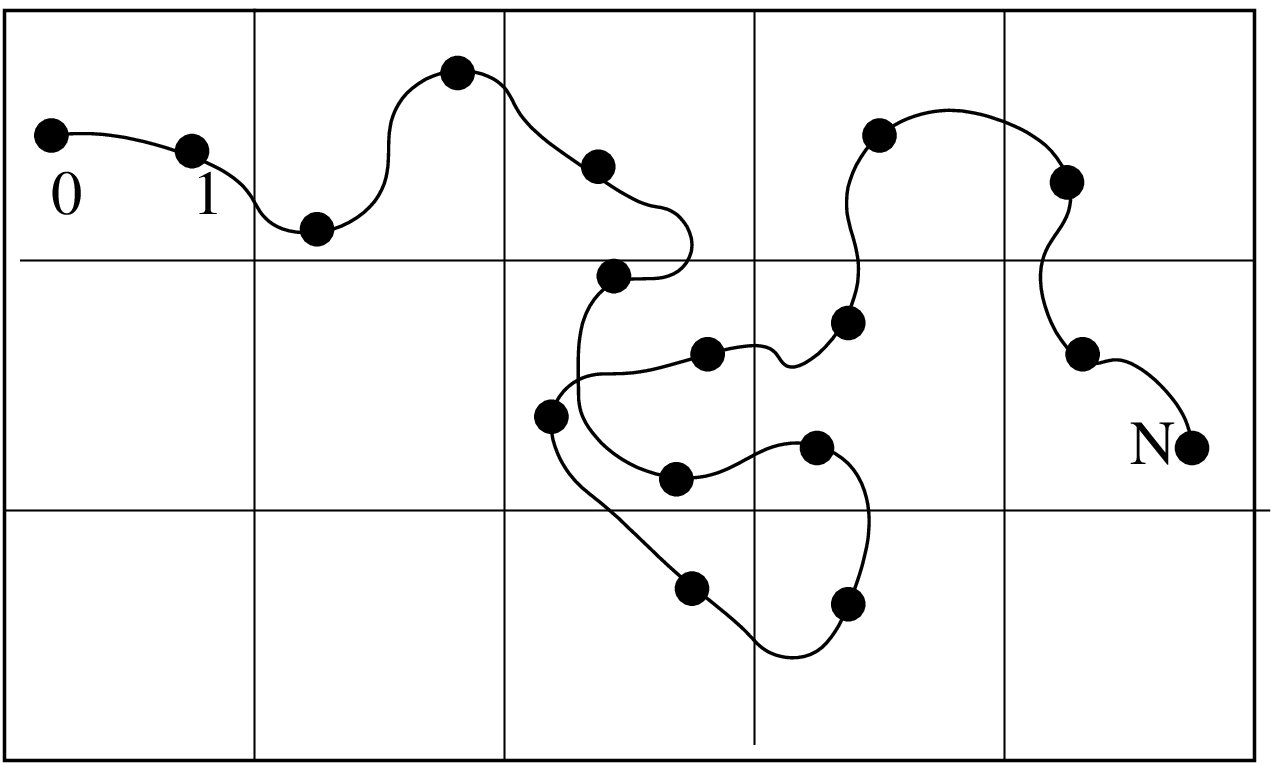}
    \caption{A chain of reptons in a lattice $d=2$ embedding}  
\label{repchain}
\end{center}
\end{figure}
A sketch of the chain in the lattice is given in Fig. \ref{repchain}.

One might worry whether the occupation of a cell by two or more reptons should be
discouraged by an energetic penalty. Of course some kind of repulsive effect is 
present due to excluded volume effect. But we must realize that the reptons
contain many monomers in a loose structure. Thus they can easily interpenetrate
each other and the error, by allowing more than two reptons unpenalized in the same cell,
is therefore less severe than on the monomer level.  In this spirit the neglect of 
self-avoidance is natural. It is not difficult to exclude more than 2 {\it successive} reptons
in a cell. That does not treat, however, the 
full problem of avoidance of more than two reptons in a cell, which implies also
interactions between remote parts of the chain. 

So the condition that the chain must form a tube is the
only restriction on the motion of the reptons. In a simulation of the
Master Equation, one moves the reptons one by one in a random fashion. In reality the 
reptons move simultaneously, but that does not give rise to a large difference 
as long as collective motion of the reptons (sliding of whole segments) is not important. 
It would mean that we shift from a stochastic motion of the reptons to organized 
motion, which is beyond the scope of this paper.

\subsection{The Hopping Rules}\label{hop}

The motion of a repton is severely restricted by the position of its 
neighbors. Thus if repton $i$ hops from ${\bf x}_i$ to ${\bf x}'_i$ we must check whether 
the new configuration forms again a tube. It means asking whether the differences
\begin{equation} \label{b2}
{\bf y}'_i = {\bf x}'_i - {\bf x}'_{i-1} \quad \quad \quad {\rm and} \quad \quad \quad
{\bf y}'_{i+1} = {\bf x}'_{i+1} - {\bf x}'_i
\end{equation} 
are still permissible links of the tube. The jump of repton $i$
\begin{equation} \label{b3}
\Delta {\bf x}_i = {\bf x}'_i- {\bf x}_i = \Delta {\bf y}_i = {\bf y}'_i - {\bf y}_i=- \Delta {\bf y}_{i+1}
\end{equation} 
is an important aspect of the move. The last equality in (\ref{b3}) stems from the fact that the
sum ${\bf y}_i +{\bf y}_{i+1}$ does not change in a move, since it is the distance between the 
reptons $i-1$ and $i+1$. Expressing ${\bf x}'_i$ in terms of $\Delta {\bf x}_i$ gives for the 
two new links
\begin{equation} \label{b4}
{\bf y}'_i = {\bf y}_i +  \Delta {\bf x}_i \quad \quad \quad {\rm and} \quad \quad \quad
{\bf y}'_{i+1} = {\bf y}_{i+1} - \Delta {\bf x}_i
\end{equation} 
An internal repton $i$ is linked by ${\bf y}_i$ and ${\bf y}_{i+1}$ to its neighbors. 
A special case is a pair of  two opposite links, which we will call a {\it hernia}. As 
the links are either slack or taut we get the following table  of possibilities for the 
jumps.
\begin{center}
\begin{tabular}{|c|c|l|}
\hline
    &  & \\*[-2mm]
type & class & \hspace*{2cm} jump \\*[4mm]
\hline 
    &  & \\*[-2mm]
slack-slack & a) & 
$ \Delta {\bf x}_i = {\bf e}_k  $\\*[4mm]
\hline
    &  & \\*[-2mm]
\begin{tabular}{c} slack-taut \\ taut-slack \end{tabular}
 & b) &  $ \Delta {\bf x}_i = {\bf y}_{i+1} - {\bf y}_i $ \\*[4mm]
\hline
    &  & \\*[-2mm]
\begin{tabular}{c} taut-taut \\ non-hernia \end{tabular} 
& c) & $ \Delta {\bf  x}_i = {\bf y}_{i+1} - {\bf y}_i $ \\*[4mm]
\hline
   &   &  \\*[-2mm]
hernia  & d) & $\left\{ \begin{array}{l} 
\Delta {\bf x}_i = -{\bf y}_i  \\*[1mm]
\Delta {\bf x}_i = -{\bf y}_i + {\bf e}_k, \quad 
{\bf e}_k \neq \pm {\bf y}_i \\*[1mm]
\end{array} \right. $ \\*[6mm]
\hline
\end{tabular}\label{tabint}
\end{center}
\begin{description}
\item[a)] Both links are slack. \\
This allows repton $i$ to hop to any of its $2d$ 
neighboring cells, thereby changing the slack links in a pair of opposite taut links, 
which is  the creation of a hernia. See Fig. \ref{repcha}. The reverse process is a hernia
annihilation. We give these hopping rates the value $h$.
\begin{figure}[h]
\begin{center}
    \epsfxsize=10cm
    \epsffile{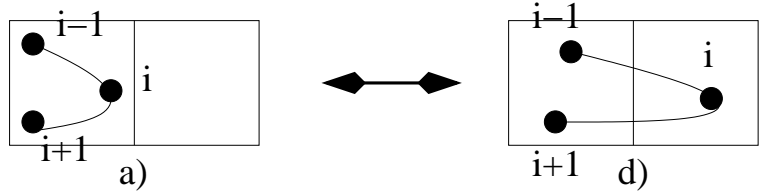}
    \caption{Hernia creation and annihilation. a) and d) refer to the class in the table}  
\label{repcha}
\end{center}
\end{figure}
\item[b)] One link is slack and one is taut. \\ 
Let link $i$ be slack and $i+1$ be taut. Then 
repton $i$ can join repton $i+1$ in its cell and the slack and taut link are 
interchanged. See Fig. \ref{repchb}. These hops will be called RD moves and 
as major hoppings, they set the time scale; so they get the rate 1.
\begin{figure}[h]
\begin{center}
    \epsfxsize=10cm
    \epsffile{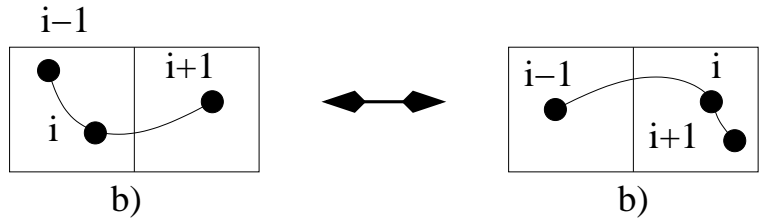}
    \caption{Rubinstein-Duke moves from class b).}  
\label{repchb}
\end{center}
\end{figure}
\item[c)] Both links are taut, but not each others opposite. \\ 
The reptons $i-1$ and $i+1$ 
then are in different cells, that both neighbor the cell of $i$. Depending on the 
circumstances repton $i$ may find another cell that again  simultaneously neighbors the 
cells of $i-1$ and $i+1$. In our (hyper) cubic lattices the only possibility is the 
interchange  of the values of link $i$ and $i+1$ (for $d>1$). See Fig. \ref{repchc}. 
These moves will be denoted as barrier crossings, with the associate rate $c$. 
\begin{figure}[h]
\begin{center}
    \epsfxsize=10cm
    \epsffile{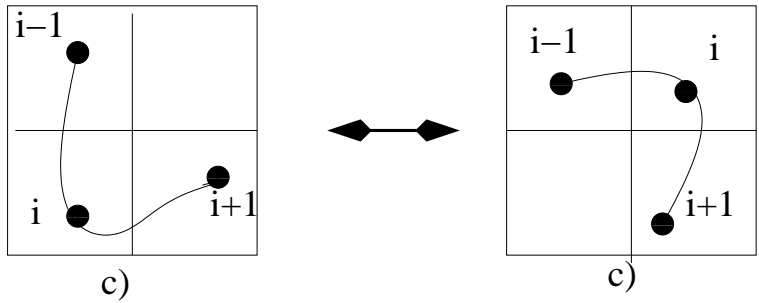}
    \caption{Barrier crossings, described in class c).}  
\label{repchc}
\end{center}
\end{figure}
\item[d)] The links are each others opposite and form a hernia. \\
Then $i-1$ and $i+1$ 
are in the same cell.  Repton $i$ then can move to this cell, which is a hernia 
annihilation; see Fig. \ref{repcha}. The other option is the migration of the hernia to any of 
the other neighbors of the cell of $i$. See Fig. \ref{repchd}. As the processes described in 
a) they get the rate $h$.
\begin{figure}[h]
\begin{center}
    \epsfxsize=10cm
    \epsffile{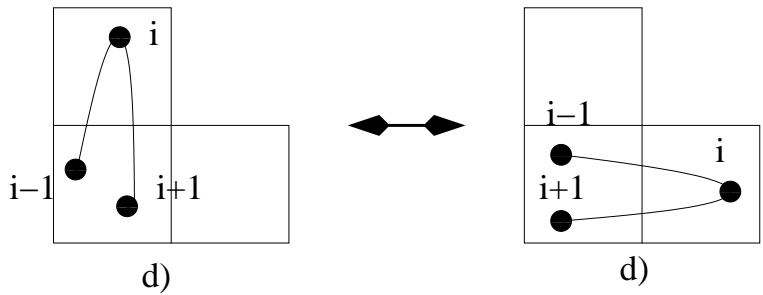}
    \caption{Hernia migrations from class d)}  
\label{repchd}
\end{center}
\end{figure}
\end{description}
In a class one may have several possibilities for the jumps; e.g. in class a) the hernia 
may be created in all $2d$ directions.

For the end reptons we get a similar list. We give it for the tail repton in the next 
table.
\begin{center}
\begin{tabular}{|c|c|l|}
\hline
    &  \\*[-2mm]
type & class & \hspace*{2cm} jump \\*[4mm]
\hline 
 &    &  \\*[-2mm]
slack & i) & 
$\Delta {\bf x}_0 = {\bf e}_k  $ \\*[4mm]
\hline
    & &  \\*[-2mm]
taut & ii) &  $\left\{ \begin{array}{l} 
\Delta {\bf x}_0 = {\bf y}_1   \\*[1mm]
\Delta {\bf x}_0 = -{\bf y}_1 + {\bf e}_k \quad {\bf e}_k \neq \pm {\bf y}_1 \\*[1mm]
\end{array} \right.$ \\*[4mm]
\hline
\end{tabular}\label{tabend}
\end{center}
\begin{description}
\item[i)] The first link is slack. \\
Then repton 0 is in the same cell as 1. Repton 0 may 
move to any of the $2d$ neighboring cells, thereby changing the first link  into a taut link.
See Fig. \ref{repchi}. Their hopping rates is 1, the same as the RD moves in class a).
\begin{figure}[h]
\begin{center}
    \epsfxsize=8cm
    \epsffile{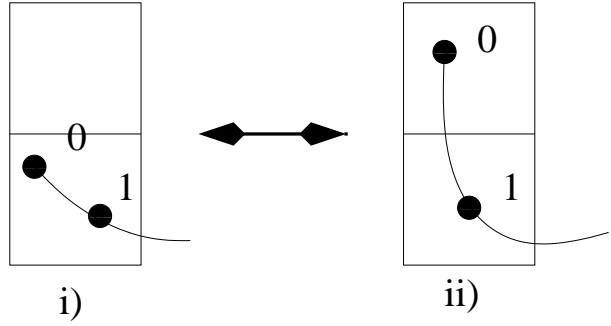}
    \caption{Rubinstein-Duke end-repton moves, class i) and ii).}  
\label{repchi}
\end{center}
\end{figure}
\item[ii)] The first link is taut. \\
Then it may join repton 1 in its cell, transforming the first link  
into a slack link (see Fig. \ref{repchi}), or move directly to the other neighboring cells 
of repton 1. These moves happen with a rate h, equal to the hernia migrations; see Fig. \ref{repchii}.
\begin{figure}[h]
\begin{center}
    \epsfxsize=10cm
    \epsffile{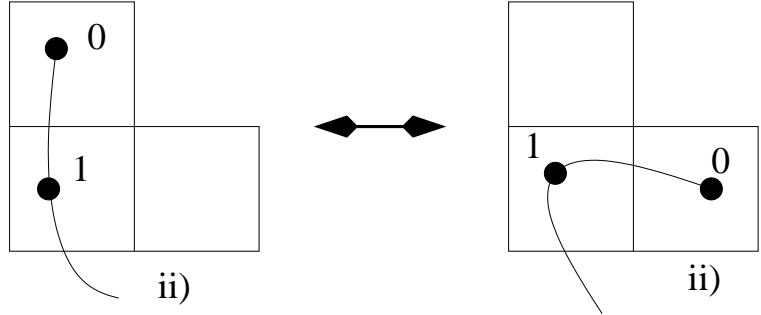}
    \caption{End-link migration in class ii), similar to hernia migration or to barrier crossing.}
\label{repchii}
\end{center}
\end{figure}
\end{description}

We denote the transition rates by the symbol $W({\bf Y}|{\bf Y'})$,
giving the transition rate from configuration ${\bf Y'}$ to ${\bf Y}$. 
We may write $W({\bf Y}|{\bf Y}')$ as a sum over the repton contributions
\begin{equation} \label{b5}
W({\bf Y} | {\bf Y}') =\sum^N_{i=0} W_i ({\bf Y} | {\bf Y}').
\end{equation} 
The first term is the transition rate of the tail repton changing the first link 
and leaving all the others the same
\begin{equation} \label{b6}
W_0  ({\bf Y} | {\bf Y}') = w_o ({\bf y}_1|{\bf y}'_1) \delta_{{\bf y}_2, {\bf y}'_2}, \cdots  
\delta_{{\bf y}_N, {\bf y}'_N}.
\end{equation} 
The next group, $W_i$, refers to the internal repton moves, which affect the two 
consecutive linkes $i-1$ and $i$
\begin{equation} \label{b7}
W_i  ({\bf Y} | {\bf Y}') = \delta_{{\bf y}_1, {\bf y}'_1}, \cdots  \delta_{{\bf y}_{i-2} , {\bf y}'_{i-2}}
w_i({\bf y}_{i-1}, {\bf y}_i | {\bf y}'_{i-1}, {\bf y}'_i) \delta_{{\bf y}_{i+1}, {\bf y}'_{i+1}}, \cdots  
\delta_{{\bf y}_N, {\bf y}'_N}.
\end{equation} 
The last term involves the head repton move changing only the last link
\begin{equation} \label{b8}
W_N ({\bf y}_N | {\bf y}'_N) = \delta_{{\bf y}_1, {\bf y}'_1}, \cdots  \delta_{{\bf y}_{N-1}, {\bf y}'_{N-1}}
w_N({\bf y}_N | {\bf y}'_N).
\end{equation}  
Note that in each term only one repton moves, all the others remain fixed. So 
conceptually the reptons move one by one. This implies the time scale for a chain update
is a factor $N$ larger than the time scale of a repton move. The former is the basic 
experimental unit and the latter the basic simulation time unit.

\subsection{The Lattice as a representation of a Gel}

The moves listed here have different implications and are used in different contexts.
Consider the embedding of the chain in a 3-dimensional cubic lattice
and imagine that the edges of the cubes form a network of obstacles and that the 
polymer can only pass through the faces of the cubes. 
This mimics the situation of a polymer dissolved in a gel. The gel forms a rigid structure
and has pores which correspond to the interior of the cells of the lattice. Here
the notion of the polymer creating a ``tube'' in which it moves is very appropriate.
The polymer can only crawl through the maze by reptation. Before a repton can hop to a 
neighboring cell, length has to be accumulated (a slack link).  
The internal moves in Fig. \ref{repchb}, which we have denoted as RD moves, are 
characteristic for reptation. They do not change the tube, but allow mass transport 
through the tube. The end reptons (Fig. \ref{repchi}) can refresh the tube configurations 
by growing and shrinking of the tube. They represent the contour-length fluctuations  (CLF).

The hernia migrations shown in Fig. \ref{repchd} look like forms of 
constraint release, but as main motion in the cage model they are pure reptations.
The reason is that in the cage model, the hernias are not counted as contributing
to the tube (contour) but seen as a form of stored length. So in the cage model stored 
length  diffuses through hernia migration. Folding and unfolding of a hernia at 
the end of the chain is then a contour-length fluctuation.
In fact we will not include the hernia migrations when discussing the RD model, 
nor include the RD moves when treating the cage model. 

The moves of Fig. \ref{repcha} and Fig. \ref{repchc} are forms
of contraint release (CR) and standardly not included in the RD model nor in the cage model.
It is a main theme of this review to discuss the interplay of CLF and CR. 
The role of the hernia creation/annihilation of Fig. \ref{repcha} is a bit ambiguous. 
They do not appear in the cage mode since it does not have the notion of slack links.
So they cannot be created and annihilation would mean the creation of a pair of slack links.
As far as the RD model is concerned they should have been included from the start, since
in a rigid structure as a gel, hernia creation and annihilation are possible without crossing
a barrier. They are left out because they spoil the mapping of Duke of a $d$-dimensional
lattice on a one-dimensional representation. Generally it has been assumed that their 
inclusion would not change the reptative character of the motion since they do not affect 
the backbone of the tube. We confirm this picture but also find that they are essential 
when allowing barrier crossing. Barrier crossing and hernia creation/annihilation have 
to cooperate in order to induce the cross-over from reptation to Rouse dynamics.

\subsection{The Driving Field}

In gel-electrophoresis the dominant influence on the polymer dynamics is the driving field.
As polymers, like DNA, are acids, the reptons become charged in a solution and an electric
field will push them through the gel. Following Duke \cite{Duke1} the driving field can be
represented by a bias in the transition rates. The bias, which has the form of a Boltzmann 
factor involving the energy difference, favors jumps in the direction of the field and discourages 
jumps against the field direction. It is wise, for symmetry reasons, to take the driving field 
in the direction of the body diagonal of the cube. Since we need this direction frequently, 
we give the body diagonal the name
\begin{equation} \label{b10}
{\bf d} = \sum_\alpha {\bf e}_\alpha.
\end{equation}
It is a vector of length $\sqrt{d}$ in a $d$-dimensional lattice.  
Then all $d$ taut links in the direction of the field are equivalent and 
this also holds for the $d$ links opposite to the field. A displacement $\Delta {\bf x}_i$
corresponds to a displacement $\Delta x_i$ in the direction of the field
\begin{equation} \label{bx}
\Delta x_i = \Delta {\bf x}_i \cdot {\bf d}.
\end{equation} 
$\Delta x_i$ is the projection of the diagonal on the displacement, 
such that a unit step $\Delta {\bf x}_i$ corresponds to $\Delta x_i = \pm 1$,
depending on whether the step is in the direction of the field or opposite.  
For repton $i$ with charge $q_i$ we take as biased transition rate 
\begin{equation} \label{b9}
w_i({\bf y}_{i-1}, {\bf y}_i | {\bf y}'_{i-1}, {\bf y}'_i) =
w^0_i({\bf y}_{i-1}, {\bf y}_i | {\bf y}'_{i-1}, {\bf y}'_i) \exp (-\epsilon q_i \,\Delta x_i/2).
\end{equation}
The superindex $0$ on $w_i$ refers to the fieldless case. The $w^0_i$ are the 
rates listed above for the various types of motion. Usually they are equal for the 
transition and its reverse. If not, it will be explicitly stated. The quantity
$\epsilon$ is a measure for the field strength. The factor 1/2 is included in the exponent 
for convenience. Such factors mean a rescaling of $\epsilon$
as measure of the field strength.
We can choose our units such that all the quantities $w_i, \epsilon, \Delta x_i$ are 
dimensionless. 

Generally we will take $\epsilon$ small for the following reason.
As mentioned several times we want $N$ to be large. Longer polymers
get hooked up around an obstacle when pulled with a finite field strength and 
little motion will result. This necessitates $\epsilon$ to be small. 
In fact Kolomeisky and Drzewinski \cite{Kolomeisky} have estimated the behavior for 
large $\epsilon$ and relatively short polymers $N<10$ and find that the probability 
distribution peaks exponentially for $U$-shaped configurations around an obstacle in 
the middle of the chain. 

So most interesting is the limiting combination
\begin{equation} \label{b11}
\lim \, \epsilon \rightarrow 0, \quad \quad \lim \, N \rightarrow \infty \quad \quad 
{\rm and} \quad \quad \epsilon N^x \quad {\rm finite}.
\end{equation}
One reckognizes this as a scaling limit. 
Depending on the combination $\epsilon N^x$ quite different behavior follows.
On the basis of simulations it is claimed that $x=1$ is the proper scaling 
power \cite{Widom,Barkema3,Barkema2}. 

Without a bias, $W_i$ is independent of the index $i$ of the repton along the chain.
Occasionally it is advantageous to have a differentiation of the hopping rates by 
given the reptons different charges $q_i$. For instance one can put a charged
repton on one end of the chain. This case is often called {\it magnetophoresis } in
contrast to {\it gel-electrophoresis} which refers to homogeneously charge chains.

A model containing simultaneously all the possible parameters would hardly be transparent.
As we will explain, not all combinations are relevant. 
In this review we will focus on phenomena in two groups of models: Repton and Cage 
models.

\subsection{Repton Models}\label{RD}

The RD model is the simplest realization of a repton model. 
As mentioned, the basic move is sketched in Fig.~\ref{repchb} and Fig.~\ref{repchi}.
The model describes reptation of a polymer dissolved in a gel. An issue is whether 
the hernia creation and annihilation, pictured in Fig.~\ref{repcha} should be included,
since these moves also do not cross any barrier. The reason for leaving them out is that 
the model simplifies considerably if only the RD moves are included. Then a projection 
of the links on the axis of the field ${\bf e}$ is possible, thereby reducing 
the link variable from $2d+1$ values to 3 values: slack, taut-forward and taut-backward. 

Although $d$ has no influence on the universal properties of the
RD model, it is an interesting parameter to play with. It controls the density of stored
length. Without a field the probability for a link to be slack equals $1/(2d+1)$. 
In this respect one does not have to restrict $d$ to an integer value. 
One can attribute to $d$ the role
of the connectivity of the lattice, as $2d$ is the number of nearest neighbors in the
(hyper)cubic lattices (that we preferentially consider). Then $d$ determines the stored
length density via the difference between extending the tube by finding a 
new cell by the head or tail and shortening the tube by retracting head or tail inside
the tube. 

This projection is not possible when hernia creation/annihilation is included. Their 
influence is one of the issues discussed in this review. It will turn out to be important
whether hernia creation/annihilation is combined with barrier crossing or not. 
It is not enriching to consider in the repton model also hernia migrations, indicated in 
Fig.~\ref{repchd}, as another perturbation. 
The reason is that a hernia migration can be seen as a succession
of a hernia annihilation and a recreation in a different direction. Thus the inclusion of
hernia migration does not add a new dynamical mode, if hernia creation/annihilation is
already present. 

In the repton models we set the hopping rates for the RD moves equal to 1. 
It defines the time scale. So we have
in addition to $\epsilon$ and $N$, two more parameters: the hernia creation/annihilation
rate $h$ and the barrier crossing rate $c$. Most interesting is to have small values
of $h$ and $c$, since we have again scaling phenomena in these parameters in combination
with the length $N$, similar to (\ref{b10}).

\subsection{Cage Models}

Cage models do not have slack links, so the link variable takes $2d$ values. The 
basic move is hernia migration, pictured in Fig. \ref{repchd} and \ref{repchii}.
We set in the cage models the hopping rate for these moves equal to 1. 
The cage model also describes reptation. The stored length is a hernia, 
which can travel along the tube (resulting from  systematic elimination of all hernias). 
As this migration leaves the inner tube invariant, the end reptons generate new
tube configurations. It leads to slow dynamics, characteristic for reptation. 
Hernia creation/annihilation is against the 
spirit of the cage model, since it requires the introduction of slack links. A new mode of 
dynamics can be added by inclusion of barrier crossings. 

It is not possible to reduce the model by a projection on the field axis. Then the 
distinction between a hernia and the combination of links around a corner (as shown in 
Fig.~\ref{repchc}) would be lost. The difference between the two is essential, since the 
former is allowed to migrate, whereas the latter can only change through a barrier crossing. 
We will show that the $d=1$ dimensional version of the cage model is exactly soluble,
but quite different in behavior from the higher dimensional embeddings 
(in contrast to the RD model).

One of the issues in the cage model is the influence of barrier crossings on the 
dynamics. It turns out to be important, changing the motion from reptation
to Rouse dynamics. 

\section{The Master Equation}\label{master}

The Master Equation for the probability $P({\bf Y}, t)$ reads in general
\begin{equation} \label{c1}
{ \partial P ({\bf Y}, t) \over \partial t} = \sum_{\bf Y'} \left [W({\bf Y} |{\bf Y}' )\, 
P ({\bf Y}', t) - W ({\bf Y}'|{\bf Y}) P({\bf Y},t) \right] \equiv {\cal M} P ({\bf Y},t)
\end{equation} 
The first term gives the {\it  gain} to the configuration ${\bf Y}$ from other states 
${\bf Y}'$ and the second term the {\it loss} that ${\bf Y}$ suffers from transitions to 
other configurations ${\bf Y}'$. The $W$'s are the transition rates which we discussed in the
previous section. The equation shows that there is no point of considering transitions
from ${\bf Y}$ to the same state ${\bf Y}$. However the matrix $M({\bf Y}', {\bf Y})$ of
the operator ${\bf M}$ contains the  diagonal elements
\begin{equation} \label{c2}
M({\bf Y}, {\bf Y}) = - \sum_{{\bf Y}' \neq {\bf Y}} W ({\bf Y}'|{\bf Y}).
\end{equation} 
The off-diagonal elements ${\cal M}$ are equal to the transition rates 
\begin{equation} \label{c3} 
M({\bf Y}', {\bf Y}) = W ({\bf Y}'|{\bf Y}).
\end{equation} 
The transitions are induced by repton hops and therefore the Master 
operator $\cal M$ is the sum of $N+1$ repton operators
\begin{equation} \label{c4}
{\cal M} = \sum^N_{i=0} {\cal M}_i, 
\end{equation}
where ${\cal M}_i$ involves the transitions $W_i $ induced by repton $i$. So
the operator ${\cal M}_i$ only affects the links ${\bf y}_i$ and ${\bf y}_{i+1}$. Seeing
the problem as a many-body system with the links as ``bodies'', the operator ${\cal M}_i$ 
is a two-body operator between nearest neighbors. This makes the problem suitable 
for a quantummechanical approach. The hamiltonian is a spin-type hamiltonian of 
nearest neighbor spin operators in a spin space of $2d+1$ components. 

\subsection{The Stationary State}\label{ss}

All initial states ultimately decay towards the stationary state. In that state the time
derivative at the left hand side of the Master Equation (\ref{c1}) vanishes. Thus it is
a solution of the equation
\begin{equation} \label{c5}
{\cal M} \, P({\bf Y}) = 0.
\end{equation} 
This set of linear homogeneous equations has a non-zero solution because
the Master operator has an eigenvalue zero, which in turn is a consequence
of conservation of probability. The matrix $M({\bf Y}', {\bf Y})$ is a so-called stochastic
matrix. As (\ref{c2}) and (\ref{c3}) show, the sum over each column vanishes. So the
left eigenvector, corresponding to the zero eigenvalue, is trivial: a constant value for 
all components. 
Without a bias the matrix of ${\cal M}$ is symmetric (a move and its reverse have the
same transition rate). Then the right eigenvector is the same as the left eigenvector
of which we just have shown that it has the same value for all the components. Thus
without a driving field, the probability distribution of the stationary state is trivial.
The problem is to find the stationary state for a finite driving field, 
as the non-trivial right eigenvector corresponding to the zero eigenvalue.

For what follows it is very convenient to employ a quantum-mechanical notation. We 
represent the stationary-state probability $P$ by $| P \rangle$ and write (\ref{c5}) as
\begin{equation} \label{c6}
{\cal M} | P \rangle = 0, \quad \quad \quad \langle U | {\cal M} = 0, 
\end{equation} 
where $\langle U |$ is the left eigenstate. Then we can write the normalization as
\begin{equation} \label{c7}
\langle U | P \rangle = 1.
\end{equation}
In terms of this Dirac notation we have the connection
\begin{equation} \label{c8}
P( {\bf Y}) = \langle {\bf Y} | P \rangle, \quad \quad \quad \langle U | {\bf Y} \rangle = 1.
\end{equation}  
Such a formulation makes transformations to a different basis easier. For instance we could
scale $\langle U |$ to a unit vector and change the normalization of $| P \rangle$ 
accordingly.

For later use we note that the individual operators ${\cal M}_i$
acting on an arbitratry distribution $P ({\bf Y})$, vanish when summed over ${\bf Y}$.
\begin{equation} \label{c9}
\langle U | {\cal M}_i | P \rangle = \sum_{\bf Y} \sum_{\bf Y'} \left [W_i ({\bf Y} |{\bf Y}' )
\, P ({\bf Y}', t) - W_i ({\bf Y}'|{\bf Y}) P({\bf Y},t) \right] = 0.
\end{equation} 

\subsection{The Gap}\label{gap}

Not only the stationary state is interesting. Also the eigenstates close to the zero
eigenvalue are of importance. The physics tells us that the probability cannot grow
indefinitely. So all the eigenvalues of the Master operator must have a negative
real part. The difference between the eigenvalue zero and the nearest non-zero is
called the {\it gap}. It measures how long it takes for a perturbation to decay towards the
stationary state. But this is precisely the inverse of the renewal time, which tells how
long the memory of a perturbation survives. The gap vanishes as a power $N^{-z}$
of the length of the chain $N$, with $z$ the dynamic exponent, which is one of the
major items of this review.

\subsection{Detailed Balance}\label{detbal}

The most important aspect of a Master Equation is the question whether it fulfils 
{\it detailed balance}. Non-symmetric Master Equations sometimes allow a detailed balance 
solution, where one can fulfil the set of equations (\ref{c5}) by the ansatz
\begin{equation} \label{c10} 
W({\bf Y} | {\bf Y}' ) \, P ({\bf Y}') = W ({\bf Y}' | {\bf Y}) \, P ({\bf Y}).
\end{equation}
The probability at ${\bf Y}'$ then follows from that at ${\bf Y}$ by multiplication of 
the ratio of the forward and backward transition rate. Thus 
one can transport the probability, through possible transitions, 
from one configuration to the whole configuration space. A conflict arises when a 
closed loop of transitions does not reproduce the same value. This happens when the 
product of the forward rates in the loop differs from the product of the backward rates.
One can show \cite{Widom} that indeed such discrepancies arise, due to the bias, for 
loops where the chain returns to the same configuration, but displaced over a distance 
in the direction of the field.

Asymmetry may also result from a projection onto a lower dimensional space.
Such asymmetries are removable since they do not violate detailed balance (since it exists
in the more detailed underlying model). This will be illustrated in specific examples.

\subsection{Contractions of the Master Equation}\label{contra}

In the discussion of the RD model in Section \ref{RD} we encountered the contraction 
of the $D^N$ configurations to $3^N$ configurations. In general contractions of the
Master Equation are possible without changing its formal structure. Suppose we want
to reduce the detailed description in terms of configurations ${\bf Y}$ to groups of
configurations, which we denote by the variable ${\bf Z}$. We assume that each state
${\bf Z}$ corresponds to a subset of the configurations space of the ${\bf Y}$ and
introduce 
\begin{equation} \label{c11}
V_Z = \sum_{{\bf Y} \in {\bf Z}} 1,
\end{equation} 
being the size of the configuration space belonging to ${\bf Z}$. The probability
on ${\bf Z}$ follows from $P ({\bf Y})$ as
\begin{equation} \label{c12}
P ({\bf Z}) = \sum_{{\bf Y} \in {\bf Z}} P ({\bf Y}).
\end{equation} 
Now we can write down a Master Equation for $P ({\bf Z},t)$ as
\begin{equation} \label{c13}
{ \partial P ({\bf Z}, t) \over \partial t} = \sum_{\bf Z'} \left [ W({\bf Z}|{\bf Z}' )\, 
P ({\bf Z}', t) - W ({\bf Z}' | {\bf Z}) \, P ({\bf Z},t) \right],
\end{equation} 
provided we define the transition rates for the ${\bf Z}$ as
\begin{equation} \label{c14}
W({\bf Z} | {\bf Z}' ) = \sum_{{\bf Y} \in {\bf Z}} \sum_{{\bf Y}' \in {\bf Z}'} 
W({\bf Y}|{\bf Y}') \, P ({\bf Y}', t) / P({\bf Z}',t).
\end{equation} 
This is a trivial relation, following from summation of the Master Equation (\ref{c1})
over all ${\bf Y} \in {\bf Z}$. In general it is also useless, because the transition rates
in ${\bf Z}$ space depend on the probability densities. However, if we know some
relation between the ratio of the two probabilities in (\ref{c14}), the expression gets
a real content. The most common case is that symmetry ensures that the
configurations ${\bf Y}$ belonging to ${\bf Z}$ all have the same probability. Then
(\ref{c12}) becomes
\begin{equation} \label{c15}
P ({\bf Z}) = V_Z \,P ({\bf Y})
\end{equation} 
and the expression for the transition rate
\begin{equation} \label{c16} 
W({\bf Z} | {\bf Z}' ) = (V_{Z'})^{-1} \sum_{{\bf Y} \in {\bf Z}} \sum_{{\bf Y}' \in {\bf Z}'}
W({\bf Y}|{\bf Y}').
\end{equation} 
Now the transition rates for the states ${\bf Z}$ are useful, since they do not depend
anymore on the probability densities and the Master Equation (\ref{c13}) can be used
to compute the probability distribution $P ({\bf Z})$. The formal connection (\ref{c16})
is an unambiguous rule to compute the transition rates in the projected space. 

The RD model is the simplest case of a possible contraction of the Master Equation to a 
smaller space. Leaving out hernia creation/annihilation and barrier crossings, the only 
distinction between the taut links is through the biases, which depend on their direction 
with respect to the field. So one may assume that the probability is the same for any of 
the $d$ up links and $d$ down links, up and down as measured with respect to the 
direction of the field. This gives the 
reduction in the configuration space from $(2d+1)^N$ to $3^N$. Consider the 3 cases; slack: 
$y_i=0$, taut-up:  $y_i=1$ and taut-down: $y_i = -1$. Then the probability $P$ can be 
simplified to:
\begin{equation} \label{c17}
P({\bf y}_1, \cdots , {\bf y}_N) = d^{-L} P(y_1, \cdots , y_N) ,
\end{equation}
where $L = \sum y^2_j$ is the number of taut links. The factor in front guarantees that
both distributions are normalized. With ${\bf Y}$ as short hand for 
$({\bf y}_1, \cdots , {\bf y}_N)$ and $Y$ for $(y_1, \cdots, y_N)$ we have
\begin{equation} \label{c18}
\sum_{\bf Y} P ({\bf Y}) = \sum_Y P (Y) = 1.
\end{equation} 
The reduction of the $2d+1$ possibilities per link to 3, shows that the embedding
dimension $d$ is of less importance. It does not totally disappear from the problem,
since the end reptons are sensitive to the embedding dimension. $d$ enters in the 
hopping rate of the end repton as a factor for some moves. This can be computed
from the prescription (\ref{c16}).  If the tail link is taut and it changes into a slack link, 
there is only one realization: by shortening the tube over one 
distance. The inverse process, the generation of a taut link from a slack link has $d$ 
possibilities for an up link and $d$ possibilities for a down link. These are 
acknowledged in the reduced description (\ref{c17}) by multiplying the transition rate
by a factor $d$. 

\subsection{Symmetrizing the Master Operator} \label{symmas}

The projection (\ref{c13}) with the transition rates (\ref{c16}) may give a welcome 
reduction in the number of configurations, but it spoils the symmetry which possibly existed
on the more detailed level. The RD model is an example of such loss of symmetry.
The rate of extension is a factor $d$ larger than the rate of retraction for an end repton.
An asymmetry between a process and its inverse is in itself not a point. The bias
by the field generally introduces an asymmetry between a move and its reverse. 
But for a weak field this field-induced asymmetry is small and easy to handle 
computationally. In general the more asymmetric the Master Operator is, the more
painful it becomes to find the solution for the stationary state. Thus it pays off to
restore symmetry when possible. The asymmetry induced by the projection can
be removed by the following transformation
\begin{equation} \label{c19}
\tilde{P} ({\bf Z},t) = (V_Z)^{-1/2} P ({\bf Z},t)
\end{equation} 
Substituting this into (\ref{c13}) yields the equation
\begin{equation} \label{c20} 
{ \partial \tilde{P} ({\bf Z}, t) \over \partial t} = \sum_{\bf Z'} \left [ \tilde{W}({\bf Z}|{\bf Z}' )\, 
\tilde{P} ({\bf Z}', t) - W ({\bf Z}' | {\bf Z}) \, \tilde{P} ({\bf Z},t) \right],
\end{equation} 
with the transition rate
\begin{equation} \label{c21}
\tilde{W} ({\bf Z} | {\bf Z}' ) = (V_{Z'} V_Z)^{-1/2} \sum_{{\bf Y} \in {\bf Z}} \sum_{{\bf Y}' \in {\bf Z}'}
W({\bf Y}|{\bf Y}').
\end{equation} 
Note that the loss term in (\ref{c20}) contains the old $W ({\bf Z}' | {\bf Z})$. 
Now if $W({\bf Y}|{\bf Y}')$ is symmetric in ${\bf Y}$ and ${\bf Y'}$ the transformed
$\tilde{W} ({\bf Z} | {\bf Z}' )$ is also symmetric as (\ref{c21}) shows. 

It is sometimes convenient to formulate this transformation in quantummechanical terms.
The (\ref{c19}) reads
\begin{equation} \label{c22}
| \tilde{P} \rangle = {\cal T} | P \rangle
\end{equation} 
and the transformation of Master Operator 
\begin{equation} \label{c23}
\tilde{\cal M} = {\cal T} {\cal M} {\cal T}^{-1}.
\end{equation} 
The left eigenstate $\langle U |$ transforms as
\begin{equation} \label{c24}
\langle \tilde{U} | = \langle U | {\cal T}^{-1},
\end{equation} 
such that the normalization (\ref{c7}) reads again
\begin{equation} \label{c25}
\langle \tilde{U} | \tilde{P} \rangle =1.
\end{equation}
Since (\ref{c23}) is a similarity transformation, the eigenvalue spectrum of
${\cal M}$ is not changed and the eigenvectors transform as $P$ viz. $U$.
 
\section{Correlations in the Chain} \label{correl}

The probability distribution $P({\bf Y})$ contains all the information on the stationary 
state, from which one can derive reduced probability distributions for parts of the chain.
The simplest is the link distribution
\begin{equation} \label{d1}
p_j ({\bf y}) = \langle U | \delta_{{\bf y},{\bf y}_j} | P \rangle .
\end{equation} 
It gives the probability that link $j$ has the value ${\bf y}$. Since the driving field
is the only vector which breaks the orientational symmetry of the lattice, there is 
still rotational symmetry around the field direction. So this probability will only
depend on the component $y$ of ${\bf y}$ in the direction of the field, which we
define like the projection (\ref{bx}) of the  displacements
\begin{equation} \label{dx}
y_i = {\bf y}_i \cdot {\bf d},
\end{equation} 
such that $y_i = 0$ for the slack link and $y_i = \pm 1$ depending whether 
the taut link is in the direction of the field or opposite to it. So one has three values
\begin{itemize}
\item for a link to be slack: $p^0_j =p_j ({\bf 0}) $,
\item for a link to be taut and in the direction of the field: 
$p^+_j = d p_j ({\bf e}_\alpha)$, 
\item for a link to be taut and opposite to the field: $ p^-_j = d p_j (-{\bf e}_\alpha) $.
\end{itemize}
The three probabilities are normalized by definition
\begin{equation} \label{d2}
p^0_j + p^+_j + p^-_j = 1.
\end{equation} 

The next reduced probability is the two link probability 
\begin{equation} \label{d3}
p_{j,j'}({\bf y},{\bf y}') =  \langle U | \delta_{{\bf y},{\bf y}_j} 
\delta_{{\bf y}',{\bf y}_{j'}} | P \rangle.
\end{equation} 
In the RD model this has again only $3 \times 3$ different values but in the general case 
one has to work with $(2d+1)^2$ values with some relations following from rotational 
symmetry around the field direction.  Higher order correlations will not be considered 
in this review.

We will frequently use transposition symmetry for symmetric chains where all the reptons
have the same charge. Then the probability distribution is invariant for interchange of
head and tail, leading to the property
\begin{equation} \label{d4}
P ({\bf y}_1, \cdots, {\bf y}_N) = P (-{\bf y}_N, \cdots, -{\bf y}_1)
\end{equation} 
and implying for the link probabilities
\begin{equation} \label{d5}
p^0_j = p^0_{N+1-j}, \quad \quad \quad p^+_j = p^-_{N+1-j}. 
\end{equation} 

\subsection{Local Orientation and Stored Length}\label{locos}

The average local orientation of the chain is defined as
\begin{equation} \label{d6}
\langle {\bf y}_j \rangle = \sum_{\bf y} p_j ({\bf y}) {\bf y} = 
\langle U | {\bf y}_j | P \rangle.
\end{equation}
This vectorial quantity has to point in the direction ${\bf d}$ of the driving field. 
Rather than using $\langle {\bf y}_j \rangle$ we consider the scalar $\langle y_i \rangle $
defined in (\ref{dx}). Using the above listing of the cases we get
\begin{equation} \label{d7}
\langle {\bf y}_j \rangle = [ p^+_j - p^-_j ] {\bf d}/d \quad \quad {\rm or} \quad \quad 
\langle y_j \rangle = p^+_j - p^-_j.
\end{equation}
$\langle y_j \rangle$ is not necessarily positive, as one sees for symmetric 
chains. Transposition symmetry implies according to (\ref{d5}) and (\ref{d6})
\begin{equation} \label{d8}
\langle {\bf y}_j \rangle = - \langle {\bf y}_{N+1-j} \rangle,
\end{equation}
showing that if the links are positively oriented at the head (which will turn out to be 
the case), they are negatively oriented at the tail.

The slack component $p^0_j$ gives the local density of stored length.

\subsection{Local Velocities}

The velocity of repton $i$ in a configuration ${\bf Y}$ is the product of the 
displacement and the hopping rate
\begin{equation} \label{d9}
 \langle {\bf v}_i | {\bf Y} \rangle = \sum_{{\bf Y}'} \Delta {\bf x}_i \, 
w_i ({\bf y}'_i, {\bf y}'_{i+1} | {\bf y}_i, {\bf y}_{i+1}) \equiv
{\bf v}_i ({\bf y}_i, {\bf y}_{i+1}).
\end{equation} 

The component along the contour, the curvilinear velocity, is obtained by taking
the inner product with the local direction of the chain, for which we take the vector
$\bf{y}_i + {\bf y}'_i$
\begin{equation} \label{dy}
\langle {\bf J}_i | {\bf Y} \rangle = \langle {\bf v}_i | {\bf Y} \rangle \cdot ({\bf y}'_i + {\bf y}_i).
\end{equation} 
This is a preliminary definition, the more general one is given in (\ref{d26}).
Clearly the projection in (\ref{dy}) works in the RD model, where for any move one of the 
${\bf y}'_i$ and ${\bf y}_i$ is zero. The non-zero value gives the direction of the contour.

Finally, we occasionally need the local current, which is the local velocity
multiplied by the charge of the repton
\begin{equation} \label{d10}
\langle {\bf I}_i | {\bf Y} \rangle = q_i \langle {\bf v}_i | {\bf Y} \rangle. 
\end{equation} 
Each of these velocities plays a role in the analysis to come.

\subsection{Drift Velocity and Diffusion Constant}\label{drift}

Driving the chain with a field results in a constant drift velocity ${\bf v}_d$ in
the stationary state. 
The average local velocity is obtained from the probability $P ({\bf Y})$ as
\begin{equation} \label{d11}
\langle {\bf v}_i \rangle = \sum_{\bf Y} \langle {\bf v}_i | {\bf Y} \rangle 
\langle {\bf Y} | P \rangle = \langle {\bf v}_i | P \rangle.
\end{equation} 
In the stationary state they must be the same for all reptons, otherwise the integrity
of the chain would be disrupted. To see this directy from the Master Equation 
we use the identity
\begin{equation} \label{d12}
\sum_{\bf Y,Y'} {\bf y}'_i \, W({\bf Y'}|{\bf Y}) P({\bf Y})= 
\sum_{\bf Y,Y'} {\bf y}_i \, W({\bf Y}|{\bf Y'}) P({\bf Y'}),
\end{equation}  
which follows by interchanging the summations over ${\bf Y}$ and ${\bf Y'}$ in one 
of the sides of the equation. Then use the stationary state Master Equation for the
the summation over ${\bf Y'}$ in the right hand side, yielding
\begin{equation} \label{d13}
\sum_{\bf Y,Y'} ({\bf y}'_i -{\bf y}_i) \, W({\bf Y'}|{\bf Y}) P({\bf Y})= 0.
\end{equation} 
The difference ${\bf y}'_i -{\bf y}_i$ can be written (with (\ref{b2})) as  
\begin{equation} \label{d14}
{\bf y}'_i -{\bf y}_i = ({\bf x}'_i - {\bf x}'_{i-1}) - ({\bf x}_i - {\bf x}_{i-1}) =  
\Delta {\bf x}_i - \Delta {\bf x}_{i-1}.
\end{equation} 
Inserting this into (\ref{d13}), the first $\Delta {\bf x}_i$ gives the average velocity 
$\langle {\bf v}_i \rangle$ and the second $\Delta {\bf x}_{i-1}$  gives $\langle {\bf v}_{i-1} \rangle$, 
leading to the relation 
\begin{equation} \label{d17}
\langle {\bf v}_i \rangle = \langle {\bf v}_{i-1} \rangle, 
\end{equation} 
showing that the local velocity is constant along the chain. 

In practice it is an important check on the accuracy of a calculation to compute the individual 
drift velocities of the reptons from the correlation functions and to see whether they are
constant as function of the position. They follow with (\ref{d9}) as
\begin{equation} \label{d19}
\langle {\bf v}_i \rangle  = \sum_{{\bf y}_i,{\bf y}_{i+1}} p_{i,i+1} ({\bf y}_i, {\bf y}_{i+1}) 
{\bf v}_i ({\bf y}_i, {\bf y}_{i+1}).
\end{equation} 
Preferably the drift velocity is calculated as the average over the chain 
\begin{equation} \label{d18}
 {\bf v}_d = \sum^N_{i=0} \langle {\bf v}_i | {\bf Y} \rangle/(N+1),
\end{equation} 
because it is more accurate than the values of the local drift velocities.
Since the drift velocity is always in the direction of the field, we often use its 
magnitude $v_d$
\begin{equation} \label{d21}
{\bf v}_d = v_d \, {\bf d}/\sqrt{d}.
\end{equation}
The zero-field diffusion constant $D$ follows from the Einstein relation
\begin{equation} \label{d22}
D = {1 \over N } \,\left( {\partial v_d \over \partial \epsilon }\right)_{\epsilon=0}, 
\end{equation}
which can be shown \cite{vanLeeuwen1} to hold for the RD model. We have dropped the 
standard factor $k_B T$ in the definition, since we work consistently in dimensionless units.
As the drift velocity is the more general quantity than the zero-field diffusion constant, 
we take it for granted that it includes the diffusion constant.

\subsection{Curvilinear Velocity}\label{curvel}

We obtained a relation for the drift velocity by multiplying the Master Equation 
with the link function ${\bf y}_j$. If we multiply with  $y^2_j$, which is 1 for a 
taut link and 0 for a slack link, we get the curvilinear velocity
\begin{equation} \label{d24}
J_j  = \langle U | y^2_j \, {\cal M}_j | P \rangle.
\end{equation} 
Working out the contributions yields
\begin{equation} \label{d26}
 J_j ({\bf y}_j, {\bf y}_{j+1}) = \sum_{{\bf y}'_j,{\bf y}'_{j+1}} ({y'}^2_j - y^2_j) \, 
w_j ({\bf y}'_j, {\bf y}'_{j+1} | {\bf y}_j, {\bf y}_{j+1}).
\end{equation} 
Note that this agrees with the preliminary definition (\ref{dy}). For the RD moves 
we get $+1$ from the moves where a slack ${\bf y}_j$  changes into a taut link 
and $-1$ for the opposite process. In the first case stored length moves from 
tail to head and in the second case it moves from head to tail. 
Thus $J_j$ gives the net motion over link $j$ along the tube regardless of its 
direction in the field and therefore $J$ is called the curvilinear velocity. 
In fact the expression is also meaningful for the non-RD moves. 
A hernia creation gives a positive contribution while
the annihilation gives a negative contribution. Neither the barrier crossings nor
the hernia migrations contribute to the curvilinear velocity. This statement must
be modified for the cage model where the contour is defined differently.

We may repeat the arguments of the previous subsection and show that
the curvilinear velocity, as defined in (\ref{d24}), is constant along the chain. 
\begin{equation} \label{d25}
J_j = J_{j+1} = J_c.
\end{equation} 
For a symmetric chain, where head and tail may be interchanged, 
the curvilinear velocity obeys the relation
\begin{equation} \label{d29}
J_j = - J_{N+1-j},
\end{equation}
which shows with (\ref{d25}) that $J_c=0$ for symmetric chains. 

In terms of the two-point correlation function it becomes
\begin{equation} \label{d27}
J_j = \sum_{{\bf y}_j,{\bf y}_{j+1}} \, p_{j,j+1} ({\bf y}_j,{\bf y}_{j+1}) 
J_j ({\bf y}_j, {\bf y}_{j+1}).
\end{equation} 

\subsection{Relation between the Tail Orientation and Drift Velocity}\label{relat}

The velocities of the end reptons depend only on the probabilities of the end links. 
As illustration we give the expressions for the RD model for the drift velocity and 
the curvilinear velocity. The velocity of the tail repton reads
\begin{equation} \label{d20}
\langle {\bf v}_0 \rangle = \left[ d p^0_1 (B - B^{-1}) + p^+_1 B  - p^-_1 B^{-1} \right] {\bf d}/\sqrt{d}.
\end{equation} 
The drift velocity is in the direction ${\bf d}$ of the field, being the 
only vector in the game. $B$ is the bias on the tail repton related to the field as
\begin{equation} \label{d34}
B = \exp (-\epsilon q_0/2),
\end{equation} 
with $q_0$ the charge of the tail repton.

The curvilinear velocity of the tail repton is given by
\begin{equation} \label{d28}
J_0 = d p^0_1 \, (B+ B^{-1}) - p^+_1 B- p^-_1 B^{-1}. 
\end{equation}
So if the probabilities on the tail link states are known, both the drift velocity and the
curvilinear velocity of the whole chain are known.

We now have several relations involving the three probabilities $p^0_1, p^+_1$ and 
$p^-_1$: the normalization (\ref{d2}), the orientation of the first link (\ref{d7}), the drift 
velocity (\ref{d20}) and curvilinear velocity (\ref{d28}). We may use the first three 
to express the probabilities in terms of $\langle y_1 \rangle$ and $J_c$. The result is:
\begin{equation} \label{d30}
\left\{ \begin{array}{rcl}
p^0_1 & = & \displaystyle {1 \over 2d+1} \left[ 1 + {2J_c \over B + B^{-1}}  + 
\langle y_1 \rangle \delta \right] ,  \\*[4mm]
p^+_1 & = & \displaystyle{1 \over (2d+1)} \left[ d - {J_c \over B + B^{-1}} -
{\langle y_1 \rangle \delta \over 2}  \right] +  {\langle y_1 \rangle \over 2},  \\*[4mm]
p^-_1 & = & \displaystyle{1 \over (2d+1)} \left[ d - {J_c \over B + B^{-1}} - 
{\langle y_1 \rangle \delta \over 2} \right] - {\langle y_1 \rangle \over 2},  \\*[4mm]
\end{array} \right.
\end{equation} 
where we used the abbreviation
\begin{equation} \label{d31}
\delta = {B - B^{-1} \over B+ B^{-1}}.
\end{equation}

For symmetric chains, with $J_c=0$, the deviations of the $p$'s from their
unbiased values (the first term) are given by the orientation of the first link. 
Since $\langle y_1 \rangle < 0$ in that case, the main effect is a decrease of $p^+_1$
and an increase of $p^-_1$. In higher order (terms $\sim \delta  \langle y_1 \rangle < 0$)
there is a decrease of stored length ($p^0_1$) in favor of the taut links.

The expressions (\ref{d30}) can be used to relate the drift velocity to the curvilinear 
velocity and the orientation of the first segment. For symmetric chains only the latter
occurs. We give for the symmetric chain the drift velocity in terms of 
the orientation $\langle y_1 \rangle$ of the first link 
\begin{equation} \label{d32}
v_d = (B + B^{-1}) \left[ {2d \delta \over 2d+1} + {\langle y_1 \rangle \over 2}
\left( 1 + {2d-1 \over 2d+1} \delta^2 \right) \right].
\end{equation} 
It shows that $\langle y_1 \rangle$ has to balance the first term very carefully in order
that the drift velocity decays as  $1/N$.

\section{Linearization in the field} \label{linear}

The driving field is a parameter with which one wants to play in applications.
In gel-electrophoresis, where the polymers are driven through a gel by an electric field, 
one likes to turn up the field as high as possible in order to speed up the experiment.
Also periodically changing the direction of the field (field inversion) is a widely used 
technique to free the configurations which are stuck behind obstacles. 
This is of course a delicate dynamical problem beyond the scope of this review. 
We are concerned with the renewal time $\tau$ and the diffusion coefficient $D$. 
Both quantities have a meaning in the driven state, but mostly their values refer to
the zero field case. The renewal time is given by the gap in the spectrum of the 
fieldless Master Operator and the diffusion time is related to the mean squared 
displacement in zero field. So both can be obtained from the zero-field 
Master Operator. It is, however, much easier to derive the diffusion coefficient from 
the linear response to an infinitesimal driving field. Rather than study the drift for
small fields and then differentiate according to the Einstein relation (\ref{d22}),
we use the simplifications that result from a direct linearization in 
the field. It boils down to an expansion in the small parameter $\epsilon$, 
which is a delicate affair, because $\epsilon$ combines with $N$. So higher orders in 
$\epsilon$ not necessarily lead to smaller contributions, even if $\epsilon$ is small. 
Nevertheless one can always find a regime where also $\epsilon N^x$  (see (\ref{b11}))
is small and for the zero-field diffusion coefficient it suffices to consider this regime. 
For the DMRG method that we are going to use, a finite field is no obstacle. But the
larger the field the more non-hermitian the Master Operator becomes. This gives a
limitation to the convergence of the method (see \cite{Carlon3} for applying DMRG 
to non-hermitian operators).

In this section we derive the formulae which facilitate the calculation of the diffusion 
coefficient. First we show that the charge distribution of the reptons 
can be eliminated in that regime. We consider mainly two charge distributions: the uniform 
one (gel-electrophoresis) and the one where only the head repton has a charge 
(magnetophoresis). Gel-electrophoresis is the more common case with many experimental
applications. The name magnetophoresis has been coined by Barkema and Sch\"utz 
\cite{Barkema1} when they introduced this model, since the envisioned experimental 
realization uses a magnetic bead and a magnetic field to pull the head repton.
After the linearization of the Master Equation we give the expressions for the 
diffusion coefficient (drift velocity).

Thus we expand the Master Equation (\ref{c5}) in powers of $\epsilon$ \cite{Drzewinski2}
\begin{equation} \label{e1}
{\cal M} = {\cal M}^0 + \epsilon {\cal M}^1 + \cdots, \quad \quad \quad P({\bf Y} ) = 
P_0 ({\bf Y}) + \epsilon P_1 ({\bf Y}) + \cdots , 
\end{equation}  
leading to the lowest order  equation
\begin{equation} \label{e2}
{\cal M}^0 P_0 = 0
\end{equation} 
and the first order equation
\begin{equation} \label{e3}
{\cal M}^0 P_1  + {\cal M}^1 P_0 = 0.
\end{equation}
The zeroth order equation (\ref{e2}) is fulfilled through detailed balance
\begin{equation} \label{e4}
W^0 ({\bf Y}|{\bf Y}') P_0 ({\bf Y}') = W^0 ({\bf Y}'|{\bf Y}) P_0 ({\bf Y}).
\end{equation}  
In most cases, where the fieldless $W^0$ equals its reversed value, the equation  
is trivially satisfied by a constant distribution 
\begin{equation} \label{e5}
P_0 ({\bf Y}) = (2d+1)^{-N}.
\end{equation} 
The value of the constant is determined by the normalization  (\ref{c7}). 
The cases where the asymmetry is a consequence of the projection of a
symmetric model the situation can be handled as in discussed in Section \ref{symmas}.

Here we are concerned with the solution of (\ref{e3}) and the properties of $P_1 ({\bf Y})$.
For the input for (\ref{e3}) we have to take the derivative of the transition probabilities 
with respect to $\epsilon$ and then set $\epsilon = 0$. We can relate it to the 
definition (\ref{d10}) of the (charge) current
\begin{equation} \label{e6}
{\cal M}^1 P_0 = - \sum^N_{i=0} \, (q_i/2) \,\sum_{\bf Y'} \left[ \Delta  x_i 
W^0_i ({\bf Y} | {\bf Y}') \, P_0 ({\bf Y}')  +  \Delta  x_i 
W^0_i ({\bf Y}' | {\bf Y}) \, P_0 ({\bf Y}) \right].
\end{equation} 
Using detailed balance (\ref{e4}) the two terms are equal and we obtain
\begin{equation} \label{e7}
{\cal M}^1 P_0 = - \sum^N_{i=0} \, q_i \, \langle v^0_i | {\bf Y} \rangle \, P_0 ({\bf Y}) = 
- \langle I^0 | {\bf Y} \rangle \, P_0 ({\bf Y}).
\end{equation} 
The right hand side contains the component of the (electric) current in the 
field direction. It has a superscript 0 since we take the zeroth order (field free) 
transition rate. Combining (\ref{e7}) and (\ref{e3}) we have to solve
\begin{equation} \label{e7a}
{\cal M}^0 P_1 = \langle I^0 | {\bf Y} \rangle \, P_0 ({\bf Y}),
\end{equation}
which is still general. 

A special case is Magnetophoresis, which has all charges zero 
except a unit charge on the head repton. It leads to the expression for the current
\begin{equation} \label{e8}
{\cal M}^1 \, P^{MP}_0  = - \langle  v^0_N | {\bf Y} \rangle \, P_0 ({\bf Y})
\end{equation}
and consequently to the problem
\begin{equation} \label{e8a}
 {\cal M}^0 P^{MP}_1 =\langle  v^0_N | {\bf Y} \rangle \, P_0 ({\bf Y}).
\end{equation}
The idea is to relate the $P_1$ of an arbitrary charge distribution to $P^{MP}_1$. 
We make the ansatz for $P_1 ({\bf Y}) $
\begin{equation} \label{e10}
P_1 ({\bf Y}) = - \sum^N_{j=1} \left( \sum^{j-1}_{i=0} q_i \right) y_j P_0 ({\bf Y}) + 
Q \, P^{MP}_1 ({\bf Y})
\end{equation} 
and show that it satisfies (\ref{e7}). Here $Q$ is the total charge of the chain
\begin{equation} \label{e11}
Q = \sum^N_{i=0} q_i.
\end{equation}
First we use the relation 
\begin{equation} \label{e9}
{\cal M}^0 \, y_j \, P_0 ({\bf Y})= ({\cal M}^0_{j-1} + {\cal M}^0_j )\, y_j \, P_0 ({\bf Y}) 
= \langle (v^0_j - v^0_{j-1}) | {\bf Y} \rangle \, P_0 ({\bf Y}),
\end{equation} 
which can be derived in the same way as in Section \ref{drift}.
With this relation the first term in (\ref{e10}) can be rewritten as
\begin{equation} \label{e10a}
{\cal M}^0 \, \sum^N_{j=1} \left( \sum^{j-1}_{i=0} q_i \right) y_j P_0 ({\bf Y}) 
= \sum_{0 \leq i < j \leq N} q_i  \langle (v^0_j - v^0_{j-1}) | {\bf Y} \rangle \, 
P_0 ({\bf Y}).
\end{equation} 
Rearranging the terms in the double sum gives
\begin{equation} \label{e10b}
\sum_{0 \leq i < j \leq N} q_i  \langle (v^0_j - v^0_{j-1}) | {\bf Y} \rangle 
= Q \langle v^0_N  | {\bf Y} \rangle - \sum^N_{i=0}  q_i \langle v^0_i  | {\bf Y} \rangle . 
\end{equation} 
Then we substitute this into (\ref{e7a}) and get
\begin{equation} \label{e10c}
{\cal M}^0 P_1 = - Q \langle v^0_N  | {\bf Y}\langle + 
\langle I^0 | {\bf Y} \rangle \, P_0 ({\bf Y}) + Q {\cal M}^0 P^{MP}_1 ({\bf Y})
\end{equation} 
Now equation (\ref{e8a}) shows that the first and last term cancel, such that (\ref{e10c})
reduces to (\ref{e7a}), which proves that the ansatz (\ref{e10}) is correct.

By (\ref{e10}) the probability distribution $P_1 ({\bf Y})$  of an arbitrary charge 
distribution $q_j$ is related to the probability distribution $P^{MP}_1 ({\bf Y})$ of the 
magnetophoresis model. The amazing aspect of (\ref{e10}) is that it is not an average
relation, but a detailed relation holding for every configuration ${\bf Y}$.

Relation (\ref{e10}) shows that the probability distributions for various charge 
distributions are quite different. So it is not obvious that the drift velocity depends 
only on the total charge $Q$. To demonstrate this, we observe that generally there is 
no contribution to zeroth order: no driving field, no drift. To first order 
we get two terms in the expression (\ref{d12})
\begin{equation} \label{e12}
\langle {\bf v}^1_i \rangle =   \langle U | {\bf y}_i \, 
\left[{\cal M}^0_i | P_1\rangle  + {\cal M}^1_i | P_0 \rangle \right].
\end{equation} 
For the MP model the second term is absent, as there is no charge on  repton $i$. 
But for the comparison we have to consider a chain with the total charge $Q$ on the last 
repton. Thus 
\begin{equation} \label{e13}
\langle {\bf v}^1_i \rangle^{MP} = \langle U | {\bf y}_i \, {\cal M}^0_i | Q \, P^{MP}_1 \rangle. 
\end{equation} 
The difference between the general expression (\ref{e12}) and the special value 
(\ref{e13}) has two contributions: the second term in (\ref{e12}) and,  in the first term,
 the deviation of $P_1({\bf y})$ from $Q P^{MP}_1({\bf y})$.  The latter works out with
(\ref{e10}) and (\ref{e9}) to be
\begin{equation} \label{e14}
{\cal M}^0_i \, [ P_1 ({\bf Y}) - Q P^{MP}_1 ({\bf Y}) ] = 
q_i \langle v^0_i | {\bf Y} \rangle P_0 ({\bf Y}),
\end{equation} 
while the former is given by (\ref{e7}), being just the opposite. So they compensate
and the drift velocity of an arbitrarily charged chain is the same  as of the chain 
with the total charge on the head repton. 

Next we apply the linearization to the calculation of the zero-field diffusion coefficient.
We use (\ref{d18}) for the drift velocity and note that there is no zero order in 
$\epsilon$, because without a field there is no drift. The expression (\ref{d9}) 
for the velocities ${\bf v}_i (Y) $ can easily be expanded in powers of $\epsilon$,
since they contain only the transition rates.  
Thus for the first order in the drift one has
\begin{equation} \label{e15}
{\bf v}_d \simeq \epsilon \left[ \langle \sum^N_{i=0} {\bf v}^0_i  | P_1 \rangle + 
\langle \sum^N_{i=0} {\bf v}^1_i | P_0 \rangle \right] /(N+1).
\end{equation} 
Then $D$ follows from the Einstein relation (\ref{d22}),
which also is independent of the charge distribution. It better be since the diffusion
coefficient is essentially a zero field property and for zero field the charge 
distribution plays no role.

The real power of relation (\ref{e10}) is demonstrated by the comparison of the local
orientation of the electrophoresis and magnetophoresis cases. Let us look first to 
the density of the slack links $p^0_j$ defined in (\ref{d1}). For a uniform charge
distribution the density of slack links is invariant for the reversal of the driving
field. Thus the value will be $1/(2d+1)$ plus order $\epsilon^2$ in the electrophoresis
case. So there is no first order contribution to $p^0_j$ and (\ref{e10}) shows that 
the same holds for magnetophoresis. Actually it is interesting to derive this directly
from the correlation properties of the magnetophoresis case. The contributions to the 
curvilinear velocity $J_j$ can be spelled out in terms of the link correlations as 
\begin{equation} \label{e17}
J_j = p_{j,j+1} (0,1) + p_{j,j+1} (0,-1) - p_{j,j+1} (1,0) - p_{j,j+1} (-1,0).
\end{equation} 
Using the normalization relations 
\begin{equation} \label{e18}
\left\{ \begin{array}{rcl}
p^0_j & =  & p_{j,j+1} (0,1) + p_{j,j+1}(0,0) +  p_{j,j+1} (0,-1), \\*[2mm]
p^0_{j+1} & = &  p_{j,j+1} (1,0) + p_{j,j+1}(0,0) + p_{j,j+1} (-1,0),
\end{array} \right.
\end{equation} 
we get
\begin{equation} \label{e19}
J_j = p^0_j - p^0_{j+1}.
\end{equation} 
As $J_j = J_c$ is constant along the chain, we find that $p^0_j$ is linearly increasing
\begin{equation} \label{e20}
p^0_j = p^0_1 - (j-1)J_c \quad \quad {\rm in \, \, particular} \quad \quad 
p^0_N = p^0_1 - (N-1) J_c.
\end{equation}
Thus the absence of first order contributions to the slack link density is equivalent
with the absence of first order contributions in the curvilinear velocity (since it is
insensitive for the direction of the driving field). But relation (\ref{e20}) is more 
interesting than this observation, because it holds for any value of $\epsilon$,
which we will use in Section \ref{localor}.

\begin{figure}[h]
\begin{center}
    \epsfxsize=12cm
    \epsffile{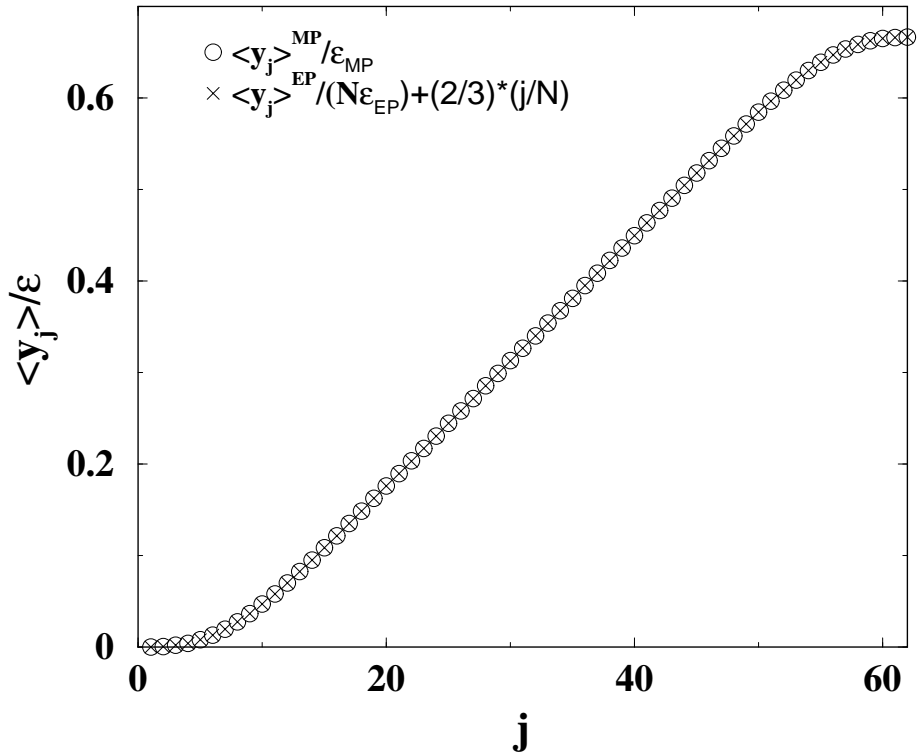}
    \caption{The orientation $\langle y_j \rangle $ for a link $j$ in first order in the field. 
Shown are the curves for the magnetophoresis case (crosses) and the corresponding points for the 
electrophoresis case (circles). The vales of $d$ is set to 1.}
\label{corfigMP}
\end{center}
\end{figure}

Comparing the local orientation for both systems we note that for zero field all
directions are equally present and no orientation appears. In first order we get from 
(\ref{e8})
\begin{equation} \label{e21}
\langle {\bf y}_j \rangle = - \epsilon  \left( \sum^{j-1}_{i=0} q_i \right) 
\langle {\bf y}_j y_j \rangle_0 +  Q \, \langle {\bf y}_j \rangle^{MP}.
\end{equation}
Here the average $\langle \rangle_0$ is performed with the probability distribution $P_0$.
This expression relates the orientation of any charge distribution (to linear order in 
$\epsilon$) to that of the orientation of the MP model. The latter is relatively dull, 
since the orientation has to be transmitted from the head to the tail. So it decreases 
from full orientation at the head to practically zero at the tail. In Fig. \ref{corfigMP}
we show the orientation in the magnetophoresis case based on a
calculation for the RD model, together with the equivalent electrophoresis curve. 
The agreement of the two curves,
calculated by independent DMRG solutions (to come), is perfect.

\begin{figure}[h]
\begin{center}
    \epsfxsize=12cm
    \epsffile{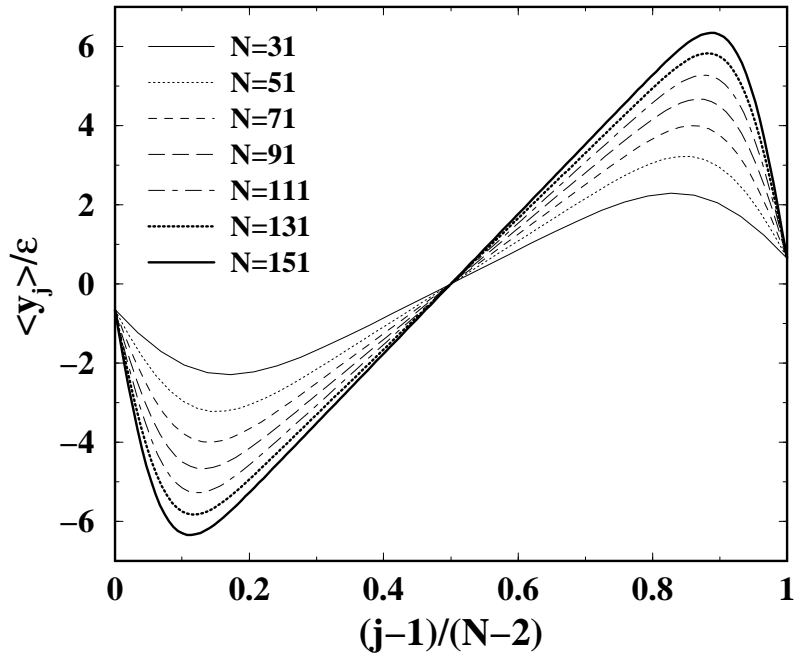}
    \caption{A collection of curves, for various values of $N$, for the orientation 
$\langle y_j \rangle $ of a link $j$ to first order in the field.
The curves refer to the electrophoresis case for $d=1$}
\label{corfigEP}
\end{center}
\end{figure}

However (\ref{e21}) would be  misleading if one were to conclude that therefore all orientation
curves are simple. In the case of an homogeneous charge distribution, the first term 
practically cancels the MP contribution and a subtile orientation effect results for the 
RD model. In Fig.~\ref{corfigEP} the first order curves are shown for the electrophoresis case.
The electrophoresis orientation has an inversion 
symmetry under transposition of head and tail, while the magnetophoresis curve is 
monotonous. 
We can go back and forth between electrophoresis and magnetophoresis to extract 
the information wherever it is easier to obtain. For example the electrophoresis curve
shows that the main effect is a linear increase from tail to head. Translated in terms
of chain shapes this means a U-form, indicating that there is a tendency to bend around
an obstacle. At the ends the effect vanishes through the fluctuations of entering
and leaving of fresh links. The end zones are estimated to be of the order $\sqrt{N}$. 
The shape of the curve shows that its calculation is far from trivial. In the 
magnetophoresis case these properties are totally obscured by the dominantly positive 
orientation. 

\section{Exact Results}\label{exact}

As mentioned in the introduction, few exact results exist, even for the simplest 
versions of the hopping rules. The systems that can be treated analytically are 
chains embedded in a one-dimensional lattice. They are soluble because the 
one-dimensional lattice restricts the possible moves, such that they can be mapped on
an exactly solvable traffic model: the asymmetric exclusion model. This model has a 
finite set of sites on a line. The sites can be occupied or not by a particle. The
particles exclude each other and they hop with different rates to the left and the
right. On the average particles are injected at the tail and extracted at the head 
side. This model can be solved by the matrix product expansion \cite{Derrida1,Derrida2}.
The essential ingredient, which makes the models soluble, is that the degrees
of freedom can be in two states: occupied or empty. 
The general case has been analyzed by Sasamoto \cite{Sasamoto} and 
independently by Blythe et al. \cite{Blythe}. 

Another case which admits an exact solution, independent of the embedding 
dimension, is the somewhat academic periodic chain.
It was treated in the idea that boundary conditions did not matter
too much, as is true for bulk phenomena. For reptation the boundary is however the
essence of the problem. Nevertheless the periodic chain yields the exact diffusion
coefficient for vanishing field. After an excursion to the periodic chain,
we list the one-dimensional models that can be viewed as traffic problems.  

 \subsection{The diffusion coefficient of the RD model}\label{diff}

In a periodic chain we have ${\bf y}_j = {\bf y}_{N+j}$, which requires that repton $j+N$
makes the same moves as repton $j$. Furthermore we take all charges equal and 
we restrict ourselves to the RD model. Having only internal reptons the tube configuration
is invariant. It can be shown \cite{vanLeeuwen2} that the probability distribution of 
the stationary state satisfies a product property. It is most easily formulated in 
terms of the excess occupations 
$n_l$ of the cells. $n_l+1$ is the number of reptons in cell $l$. We start counting
from a certain cell 1 to the last cell $L$, after which the periodic image of cell 
1 appears. So we have the relation  
\begin{equation} \label{f1}
N = \sum^L_{l=1} (1 +n_l ) = L +  \sum^L_{l=1} n_l. 
\end{equation} 
The above mentioned product property then reads
\begin{equation} \label{f2}
P(n_1, \cdots , n_L) = \prod^L_{l=1} p_l^{n_l},
\end{equation} 
where $p_l$ depends on the tube configuration. This is formed by the taut links,
which we may number $s_1, \cdots, s_L$,  with $s_i$ connecting cell $i$ to $i+1$. 
$s_L$ connects cell $L$ to the periodic image of cell 1. Being taut links, the $s_i$ have 
the values $\pm 1$. The expression for $p_l$ is found to be \cite{Kooiman1}
\begin{equation} \label{f3}
p_l = {1 \over L} \sum_n \exp [ \epsilon \sum_i a^n_i s_{i+l-1} ],
\end{equation}
where the matrix $a$ is given by 
\begin{equation} \label{f4}
a^n_i = 1/2 \quad {\rm for} \quad i>n, \quad a^n_i = 0 \quad {\rm for} \quad i=n, \quad 
a^n_i = -1/2 \quad {\rm for} \quad i<n. 
\end{equation}
The normalization of the $p_l$ is such that they approach 1 in the zero-field case 
($\epsilon = 0$). Equation (\ref{f4}) shows that the dependence of the probability 
distribution on the tube configuration ${\bf S}$ is quite intricate and contains long 
range correlations.

The solution (\ref{f2}) is rigorous and holds for arbitrary field strength, but the 
periodic chains give no clue, which distribution for the 
tube configurations has to be taken, in order to perform a meaningful average over tubes.
In the open chain, tubes and occupations mutually determine each other. Here the 
tube distribution has to come from an outside source. For weak fields, a logical choice 
is the random distribution, giving to each tube the same probability, which leads to
the result
\begin{equation} \label{f5}
D \simeq {1 \over (2d+1) N^2}.  
\end{equation} 

The periodic chain is an interesting exercise in statistical physics, but it has a limited
value for polymer motion. The reason is that the periodicity introduces a host of 
conservation laws. In fact every tube configuration is an invariant. This leads formally
to an infinite renewal time. Only the occupation
of the cells of the tubes is a dynamical variable. Therefore the average over tube
configurations is arbitrary and only in the limit of weak fields, the random average can 
be related to the true distribution. This was made rigorous by Widom and Al-Lehyani
\cite{MWidom1}. Earlier Pr\"ahofer and Spohn \cite{Prahofer} arrived at the same 
conclusion by linearizing the Master Equation in the field. Their work contains also
detailed numerical information on the local orientation in the weak field limit. One 
wonders whether their proof can be simplified by applying it to the Magnetophoresis 
model and then using the general relation (\ref{e8}) to carry the conclusion to the 
RD-model. We also point out that the expression (\ref{f5}) refers to any chain with 
the same total charge.

\subsection{The one-dimensional Cage Model}

The cage model in one dimension has a very simple representation. It is a sequence of 
forward links with $y_i=1$ and backward links $y_i=-1$. Motion is possible for a pair
$1,-1$ (a hernia) which can change into a pair $-1,1$ and vice versa. In the first 
option a repton moves backwards over two lattice distances and it gets a bias $B^{-2}$. 
In the reverse process it has a bias $B^2$. Barrier crossings are impossible because 
the reptons cannot move sideways. It is natural to see 1 as a particle and $-1$ as an 
unoccupied site. The flipping of a hernia corresponds with a particle-hole interchange
or a hop of a particle.
Thus this cage model becomes equivalent with the above mentioned traffic model. 
The end reptons can change the value of the first (last) link, thereby
changing it from a 1 to a $-1$ or vice versa. In the traffic language this means 
destroying or creating a particle viz. injecting or extruding 
a particle. The rates are determined by the displacement of the end repton and either
$B$ or $B^{-1}$. Thus the traffic model has a special combination of rates, which 
puts it in the class of maximum current models. 

The interesting point of the traffic model is that it can be solved for arbitrary 
values of $B$. \cite{Drzewinski4} The drift velocity reads for large $N$
\begin{equation} \label{f6}
v_d = {1 \over 4} (B - B^{-1}).
\end{equation} 
Note that for small $\epsilon$ the drift becomes indeed proportional to $\epsilon$,
which gives a diffusion coefficient $D \sim N^{-1}$.

Not only the gap can be calculated explicitly, but the whole spectrum of the 
zero-field hamiltonian, since it is equivalent with the Heisenberg ferromagnetic spin
chain. The gap reads (for any $N$)
\begin{equation} \label{f7}
\Delta = -2[1 - \cos(\pi/N)].
\end{equation}
The gap approaches for large $N$
\begin{equation} \label{f8}
\Delta \simeq {\pi^2 \over N^2},
\end{equation}
which is characteristic for Rouse dynamics.

\subsection{The Necklace Model} \label{necklace}

A model which falls outside the types that we list in the section \ref{hop} is the 
necklace model. It was introduced by Guidoni et al. \cite{Terranova} who calculated
the curvi-linear velocity. Since the necklace model is
rather close in spirit with the RD model and as it is exactly soluble, we mention
it here. The necklace is a string of beads on a line of sites. The beads are either
neighbors or next-nearest neighbors. In the latter case there is an unoccupied site
between the beads. Beads may not occupy the same site. The integrity of the chain is
enforced be requiring that two consecutive unoccupied site are forbidden. The vacancies
represent the internal motion of the chain, which consists of interchange of 
vacancies and beads. At the ends a bead may interchange with a vacancy from outside. 
Clearly the number of beads is conserved, but not the number of vacancies, 
which may enter and leave at the ends.

The key for the exact solution is to focus on the motion of the vacancies 
rather than on the beads. With $N+1$ beads there are $N$ possible positions 
for the vacancies available, because each vacancy has to be surrounded by beads
The $N$ positions may carry a vacancy or not. The state of the chain is fully 
characterised by the distribution of the vacancies. Then view a vacancy as a 
travelling particle and the beads as room for the particle to interchange with.
The vacancies obey exactly the same rules as the particles in the traffic 
problem. The map on the traffic model has been worked out in
\cite{Drzewinski3}. It gives the diffusion coefficient and renewal time
for an arbitrary driving field. Their $N$ dependence is the same as that for reptation.

\subsection{Curvi-linear Diffusion}

A related problem, the curvilinear diffusion coefficient, has been exactly worked
out by Buhot \cite{Buhot} for the repton model. The repton model was
originally designed by Rubinstein \cite{Rubinstein} without the a driving field.
Then each link has only two possible values $y_i=0$ for a slack link and $y_i=1$ 
for a taut link. As Buhot notices, this model is equivalent with the necklace model,
with the slack links (the mobile units) corresponding to the vacancies and the 
taut links corresponding to the beads of the necklace. The slack links diffuse along 
the tube traced out in space by the chain; so their motion determines the 
curvi-linear diffusion coefficient. By studying the mean square displacement 
of the center of mass
as a function of time, starting from a sharp initial condition, Buhot could 
derive an expression for the curvilinear diffusion coefficient. This study 
complements the solution described in the section \ref{necklace} as it 
holds for an arbitrary embedding dimension $d$ (influencing only 
the density of slack links). But it is restricted to zero field, while the necklace
solution only holds for embedding dimension $d=1$ but applies to arbitrary 
field strengths. 

\subsection{A herniating one-dimensional RD model}\label{hernia1}

The RD model shows the slow diffusive behavior (\ref{e8}), proportional to $N^{-2}$, 
typical for reptation. The motion is strongly inhibited, as forward
links $+1$ and backwards links $-1$
cannot interchange. Their order (the tube configuration) can only be broken down
and rebuild by the external reptons. If one allows hernias to be created and annihilated,
the projection on the driving field is not possible. That made the dimension 
$d$ of the
embedding lattice a minor ingredient in the RD model, not influencing the universal 
behavior such as the $N^{-2}$ dependence of the diffusion. The general influence 
of hernias will be discussed later. Here we take the special case of a one-dimensional
sytem with a hernia creation and annihilation rate equal to the hopping rate. 
We discuss this case here mainly because it is a nice illustration of the general technique
of reducing the Master Equation.

Sartoni and van Leeuwen \cite{Sartoni} noted that annihilating a pair of 
opposite links and recreating them in the opposite order, boils down to the 
interchange of forward and backward links. So the obstacles, that make the 
RD model so slow, are 
effectively removed. They suggested that the system can be mapped on a  
system of two types of particles $+$ and $-$, which hop independently of
each other from link to link. The $+$ particles exclude each other, as do the 
$-$ particles. So a link can be: either empty, occupied by a $+$ or a $-$ or by both,
which we denote as $\pm$. The total number of possible occupations is $4^N$.
A $+$ particle can hop to an empty link or to a link occupied by a $-$ particle,
leading to a $\pm$ occupation. Similarly a $-$ particle can move to an empty link or
to a link occupied by a $+$ particle, also creating a $\pm$ occupation. The simplicity of
the model stems from the fact that the $+$ and $-$ systems do not interact and the
probability on the total is the product of the probabilities of the two systems, which are
an image of each other when one simultaneously reverses the driving field. 

The next step is the contraction of the two states, empty and doubly occupied, 
to a single state. The resulting Master Equation becomes useful if the 
probabilities for both states are assumed to be  equal. It turns out that this 
ansatz is in general not satisfied, but that it is correct to linear order in 
the field. So it produces correctly the gap and the diffusion constant. Since 
the detailed arguments are a bit involved they are presented in the appendix.

\section{The Density Matrix Method}\label{dmrg}

If detailed balance is not fulfilled, no general method exists to solve the Master Equation
(apart from simulating the equation.) Here we discuss a method
to deal with the Master Equation for linear chains, which is based on the analogy with
one-dimensional quantum systems. White \cite{White1} designed a technique for treating
spin hamiltonians and baptized it the Density Matrix Renormalization Group
(DMRG) method. Whereas Renormalization Group in itself is already a misnomer 
(it is no group), the word renormalization is also inappropriate for White's technique. 
Nevertheless the method has become known as the DMRG method and we go along with
that name. The general idea is to find a representation of the groundstate wavefunction
in a restricted space which is much smaller than the original configuration space. 
The game is to find in a subspace of given size, optimal basis states in which the 
hamiltonian can be expressed. Restricting oneself to a smaller subspace for calculating 
the groundstate eigenvalue of the hamiltonian is in itself not an approximation.
If one were to know the exact groundstate eigenfunction, the one-dimensional subspace 
spanned by it, would do. However, to exactly find the wavefunction in the 
$(2d+1)^N$-dimensional space, is an illusion and the best one can do is to exploit 
the knowledge from smaller systems for larger systems. 
The special point of the DMRG technique is that it 
takes advantage of the geometry of the systems. Therefore it works phenomenally
in one-dimensional systems with short-ranged interactions, but tediously in higher 
dimensions.  A one-dimensional chain can be meaningfully split into  a ``left'' and
``right'' block,  which only interact via their connection point. Also open systems give 
more accurate results than  circular systems, which have a double interaction zone
between the two parts.

All these features make the linear chains with local hopping rules, ideally suited for
treatment by the DMRG method. As we mentioned in the Introduction the Master
Operator can be seen as a spin hamiltonian for a $2d+1$ spin-component spin-system. 
The language of the DMRG method is more
comfortable in terms of hamiltonians and eigenfunctions. So we define a hamiltonian
\begin{equation} \label{x1}
{\cal H} = - {\cal M}
\end{equation} 
as an equivalent of the Master Operator. The advantage is that the stationary state
of the Master Equation is the lowest eigenstate of $\cal H$ and the excited states
correspond to the decaying states of the Master Equation. 

The stationary state of the Master Equation translates into the groundstate of ${\cal H}$.
The chain is splitted into a {\it tail block} and 
a {\it head block} at an arbitrary repton. The optimal basis is constructed in the product
space of the two blocks. Information of the head block helps to find the optimal basis 
for the tail block and vice versa. It is reached by gradually enlarging the chain, 
using the information on shorter chains for longer ones.
So one starts with a chain small enough for an exact determination of the groundstate. 
Then a reduction of the basis is performed in order to handle larger systems. 
The reduction consists of selecting a limited set of $m$ eigenfunctions of the density 
matrix as basis. In its simplest version, the density matrix of the tail block is constructed 
from the groundstate wave function of the whole chain by tracing out the variables in 
the head block. This is the main difference with standard renormalization techniques
in which the choice of states is determined by optimizing the (free) energy. The 
advantage is that the DMRG technique is a variational method to optimize the
groundstate wave function. The largest eigenvalues of the reduced density 
matrix give the best basis functions. All the eigenvalues add up to 1. 
The sum of the eigenvalues which are left out, yields the 
{\it truncation error}. It is an indicator for the quality of the approximation. The system 
itself tells what is a reliable number of states kept \cite{Kaulke}. In the most 
succesful applications the truncation error is of order $10^{-13}$.

One can also check the quality of the approximation by calculating the groundstate 
eigenvalue on the restricted basis and compare with its known value.  
In the reptating chain one should find an eigenvalue zero for the groundstate for
all lengths of the chain. Next to the truncation error this is also an indicator for the accuracy 
of the process. 
\begin{figure}[h]
\begin{center}
    \epsfxsize=12cm
    \epsffile{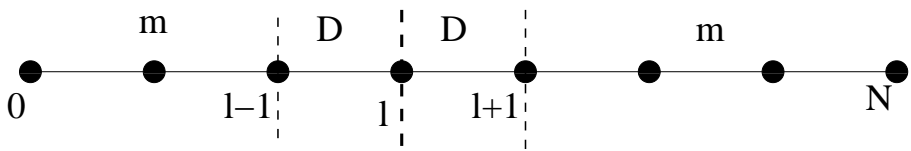}
    \caption{The division of the chain in $l$ links left and $N-l$ right.}  
\label{DMRG}
\end{center}
\end{figure}

The technique gets a major improvement by sweeping through the chain via shifting
the dividing repton. To be specific consider the situation where we have repton $l$ as
dividor and we want to shift it to repton $l+1$, see Fig. \ref{DMRG}. From
the tail block we take the last link $l$ apart. It interacts via repton $l$ with the first link $l+1$ 
of the head block. The trick is to formulate the action of the reptons inside 
the tail block in terms of a $m \times (2d+1)$ basis, where $m$ is the choice of $m$ basis 
states, representing the $(l-1)^{2d+1}$ original states for the first $l-1$ links.
$2d+1$ is the number of configuration of the link $l$ next to the dividing repton.
The same representation is chosen for the head block 
in terms of $(2d+1)$ states for link $l+1$ and $m$ states for the remaining $N-l-1$ links. 
With the full basis for the links $l$ and $l+1$, we can express the action of connecting 
repton $l$, as it works on these two links. So we have to use as basis for the total 
hamiltonian in total  $m \times (2d+1) \times (2d+1) \times m$ states. 
Next the reduced density matrix for the tail block is calculated in the $m \times (2d+1)$ 
dimensional basis and $m$ new optimal states are selected.  
The $m \times (2d+1)$ basis for the tail block is  then 
contracted to a new $m$  basis for the tail block. With the transformation matrix from
old to new basis, we find the representation of the hamiltonian in the new basis states.
Finally we combine this basis with the $2d+1$ states of link $l+1$ to form the hamiltonian 
for the new tail part and we can start the calculation with repton $l+1$ as dividor.

So we get a growing tail block at the expense of the head block. For every division the
representation of the tail block hamiltonian is kept. This is used in the backwards
sweeping phase. When the division reaches a head block so small
that $m> (N-l-1)^{2d+1}$, the head block is represented without error and the sweeping
process is reversed. Now the head blocks are successively upgraded with the help of 
the just calculated tail-block representations. It can be shown \cite{White2}
that in each step the quality of the groundstate improves. It shows up as a monotonic
approach of the calculated groundstate eigenvalue towards 0.

After an (usually small) number of sweeps, the groundstate has  converged and
one can enlarge the chain by inserting 2 new links in the middle of the chain, changing
the length from $N$ to $N+2$. Then again the first step is calculating the groundstate
for the longer chain, with the best representation for a tail and head block of $N/2$  
links, together with the full $(2d+1)^2$ states for the inserted links.  Once a groundstate
is available, reduced density matrices can be calculated and the chain can be swept
towards improving groundstates. 

In principle, all the information of the groundstate on the full basis remains available,
if all the transformation matrices are kept of an $m \times (2d+1)$ representation to an
$m$ states representation \cite{duCroo}. This is however a large load on the memory
and interesting information, such as local correlation functions, are available without all 
these data.
As the groundstate always contains two successive links in full representation, the
correlation between two successive links can directly be extracted from the groundstate. 
This allows to calculate the local orientation and local drift velocity of the dividor 
repton. The constancy of the local drift velocity along the chain is another internal check 
for the computation. 

As result of this method one obtains, starting from an exactly calculable chain,  a series
of all even-length chains up to the length one desires. Clearly the calculation becomes
slower, the longer the chain as the sweeps get longer. The most time consuming 
step is the determination of the groundstate for the full chain. The factor, which influences
the speed of the calculation dramatically, is the number $m$ of states kept. It 
determines the size of the basis,  via $m^2$, in which the groundstate calculation is
performed. So a compromise has to be drawn between one's patience, the available 
memory and the desired accuracy \cite{DMRGbook}.

With increasing $(2d+1)$ (the full number states per link), one has to shift $m$ to larger
values, of the order 50-100, for a stable calculation with great accuracy. For problems,
where the dimensional reduction of the RD model does not apply, calculations for $d=3$
(where $2d+1=7$) are not feasible, without additional speed-up tricks derived from
symmetry considerations. Mostly they apply to the linearized situation (\ref{d3}), where
one can use the full symmetry group of the lattice. In the case of a finite driving-field the
symmetry is restricted to rotations around the field direction. The eigenstates of the
hamiltonian occur in sectors dictated by the symmetry. By diligent use of these
sectors for the tail and head (and rules how to combine sectors for the tail and head
to sectors for the whole chain), one can reduce the number of states with respect to the
number $m$, that one otherwise would be needed for the same accuracy. In particular the
gap calculation profits from these symmetry considerations, because the gap belongs
to a different symmetry sector than the groundstate and inside each symmetry sector
one can use the same technique as for the groundstate to find the smallest eigenvalue.
Fortunately, as long as the truncation error remains small, the results are virtually 
exact and they are therefore ideally suited for finite size analysis, which we illustrate in the 
next section for a few examples. 

One may wonder how it is possible that longer and longer chains can still be represented
by a basis with a fixed number $m$ of states. A partial explanation follows from the
structure of the correlations in the chains. The picture that emerges (see Fig.~\ref{corfigEP})
 consists of three
zones: two short ones, of order $\sqrt{N}$, at the ends of the chain and a long one, of
order $N$, in the middle. The middle zone has little structure and is likely  well 
represented by a small basis, the end zones have a delicate structure but grow only
slowly with the length of the chain. This means that inserting new links in the middle
of the chain is a mild perturbation of the wave function.

\section{Finite-Size Analysis}\label{finitesize}

Since few analytical results exist, the asymptotic behavior must be deduced from numerical 
data for finite chains.  Finite-size analysis is a systematic tool to establish the asymptotic 
behavior. It has been developed in the theory of critical phenomena \cite{Privman}.
There the main goal is to find the precise exponent which controls the asymptotic behavior.
In polymer dynamics the situation is somewhat different, because the theoretical values 
of the exponents are well known for the models we consider here. 
The problem is to find how large the value of 
the length $N$ has to be in order that the properties exhibit the expected asymptotic 
behavior. Since the introduction of the idea of reptation by de Gennes \cite{deGennes1}, 
the theoretical estimate for the renewal time $\tau$ has been $\tau \sim N^3$. 
For the RD model this value is also found by the following argument. Renewal of the
tube configuration results from inward diffusion of fresh taut elements from the ends of 
the chain. The distance over which the links have to diffuse, for a complete 
refreshment of the chain, is of the order $N$ and their number is also of the order $N$.
By a concerted motion it would take of the order of $N^2$ of repton hops before the chain
has renewed itself. The process is however not systematic but diffusive and 
therefore the time needed is $N^4$ repton hopping times. 
Measured in chain updates rather than link updates one arrives at $\tau \sim N^3$. 
As mentioned earlier (see Section \ref{physics})  this implies for the diffusion coefficient 
$D \sim N^{-2}$.

In spite of this clear estimate, there has been a longstanding debate on the value of the
exponent, because measurements of the viscosity lead to a value $z=3.4$. De Gennes
\cite{deGennes1} considered it as one of the still open problems in polymer physics.
The measurements concern however the polymeric behavior in melts, 
which is a less clear context for reptation as compared to gel-electrophoresis.
Thus several scenarios are still open. One of them is the possibility that the discrepancies
between theory and experiment are to be blamed on finite-size corrections 
\cite{Paessens1,Carlon1}. This motivates to carefully investigate the RD model and to make 
an estimate where and how the asymptotic behavior sets in. 

The RD model has been the subject of extensive simulations, from which the asymptotic
behavior emerges \cite{Barkema3}. As mentioned in the Introduction, simulations get
lengthy, because of the steeply increasing  renewal time and therefore become inaccurate 
for large $N$. The DMRG method is able to handle intermediately long chains with a high
accuracy. Renewal time and the diffusion coefficient are calculated by separate
programs. 

The renewal time is found from the first excited state and its gap with respect to the 
groundstate. The DMRG method is geared to find the groundstate (stationary
state) of the system. So the gap state has to be made an extremal state of the system. 
The most elegant method is to find the symmetry of the gap state, which turns out
to be different from the groundstate. For instance the gap state is odd under transposition 
of the chain, while the groundstate is even. If one can restrict the search to the symmetry
sector of the gap then the DMRG method can be applied as usual. Another general trick
is to add a shift to the Master Operator (in the form of a hamiltonian ${\cal H} =- {\cal M}$ )
\begin{equation} \label{h1}
{\cal H}' = {\cal H} + \Delta | \phi_0 \rangle \langle \phi_0 |.
\end{equation} 
where $|\phi_0 \rangle$ is the (zero field) groundstate. Acting on the groundstate it 
gives an eigenvalue $\Delta$. The eigenstate of the 
first excited state is orthogonal to $| \phi_0 \rangle $ and it is also an eigenstate of
${\cal H}'$ with the same eigenvalue. By choosing $\Delta$ sufficient
large, the gap state becomes the lowest eigenvalue of ${\cal H}'$.
It is easy to target this state by the DMRG method, since the matrix of ${\cal H}'$ is 
symmetric and DMRG workes better and faster for symmetric matrices.
The drawback of this trick is that one has to target both the gap and the groundstate, 
which enlarges the necessary basis and slows down the calculation \cite{Carlon1}.

The diffusion coefficient is found from the drift velocity in a small field. As explained 
in Section \ref{linear} this can be reduced to the solution of a linear set of equations 
(\ref{e3}). In the full configuration space the number of equations grows exponentially
with the length of the chain. So exact solution is limited to short chains. However, in 
the same spirit as finding the eigenfunctions of the Master Operator, we can expand
the solution of the linear set of equations in terms of the most probable state. 
In other words the DMRG technique can be directly applied to solving the equations. 
The matrix of the equations  is symmetric. We have a set of equations of the type
\begin{equation} \label{h2}
\sum_j H_{ij} X_j = V_j,
\end{equation} 
which can be considered as the minimalization equation of the function
\begin{equation} \label{h3}
K = {1 \over 2} \sum_{i,j} X_i H_{ij} X_j - \sum_i X_i V_i.
\end{equation} 
Note that it is essential that the matrix $H$ is symmetric in (\ref{h3}) for leading 
to the minimalization equations (\ref{h2}). Since ${\cal M}^0$ in (\ref{e3}) 
refers to the fieldless case, this requirement is fulfilled. Minimalization of $K$ is most
conveniently achieved with the conjugate gradient method \cite{Numerical}. The DMRG
method consists of representing the matrix $H$ and the vector $V$ on an optimal basis
which in turn is determined with the previous solution $X$ and $V$ as input.

DRMG not only yields the drift velocity but also the  correlation functions $p_j$ and 
$p_{j,j+1}$, introduced in Section \ref{localor}. 
From these the local drift velocity can be calculated. As a check one
uses the fact that the local drift has to be the same everywhere. The most accurate value 
for the drift velocity follows from (\ref{d18}).

\subsection{The RD renewal time $\tau$}\label{renew}
 
DMRG gives gap values for a series of even length chains up to $N=100$ to $200$ . 
The standard procedure to get the power, is to plot $\log \tau(N) $ against $\log N$, 
hoping that the curve will be straight, such that an exponent can be deduced from the 
slope of the curve. This does not work very well for reptation as we will see. The following 
alternative is more accurate \cite{Carlon1}. Define an effective exponent
\begin{equation} \label{h4}
z_N = {\log \tau(N+1) - \log \tau(N-1) \over \log (N+1) - \log (N-1)}
\end{equation} 
and plot this against $N^{-1/2}$. The power $N^{-1/2}$ is for the moment a matter of 
expediency, since it gives more clear conclusions. In Section \ref{localor} we give
as reason for using this power. 
\begin{figure}[h]
\begin{center}
    \epsfxsize=12cm
    \epsffile{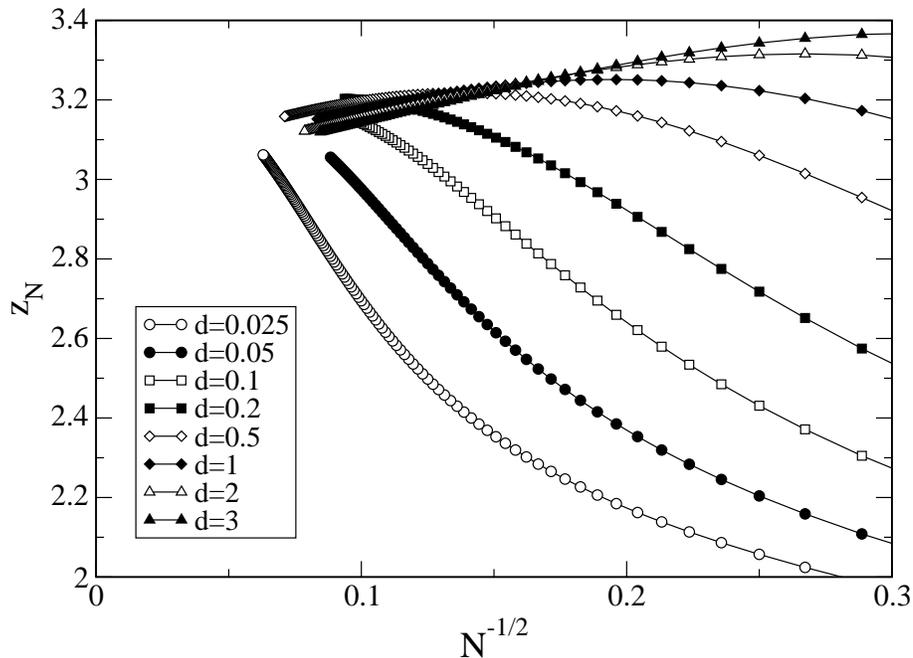}
    \caption{The effective exponent $z_N$ as function of $N^{-1/2}$ for various 
embedding  dimensions $d$.}
\label{effexp}
\end{center}
\end{figure} 

Figure \ref{effexp} shows $z_N$ 
as function of $N$ for a number of values of the embedding dimension 
$d$. If the dependence of $\tau$ on $N$ would be a pure power, the curves would be
constant. Clearly there is substantial dependence on the length $N$ of 
the chain. The points lie on smooth curves, indicating that they
are not plagued by statistical noise. Determining the value for chain lengths of the 
order of 100 reptons, together with the known asymptotic exponent, permits
to make by interpolation a reasonable estimate of the effective exponent for chain 
lengths far beyond the actual calculated values. A cubic curve fits the calculated
data as well as the exact asymptotic value of the exponent.
It is the strength of the accuracy of the DMRG calculations, that ratios of small differences
as in the expression (\ref{h4}) for $z_N$, remain smooth.
We deduce from this plot the following points:
\begin{itemize}
\item The most important feature is that the curves allow to determine the effective exponent
for a given range of $N$ values, also for ranges way beyond those computed, since in
the extrapolation to $N \rightarrow \infty$, the curves aim at the value  $z_N = 3$,
for most values of $d$. Even at very large values of $N$ the effective exponent is 
substantially larger than 3. So the discrepancy between theory and experiment can be 
partly attributed to finite-size corrections. The asymptotic value for $N \rightarrow \infty$ 
is consistent with  the theoretical estimate $z_N = 3$ for all $d$. The plateau in 
some of the curves can be misleading in assessing the asymptotic behavior if 
measurements are mainly obtained around the plateau values. In a log-log plot such
a plateau can be easily mistaken for asymptotic saturation. However,
the plateau is not several decades wide, as is often found in experiments.
\item The embedding dimension $d$ has a large influence for intermediate chains of the 
order of $50-500$ reptons (which may correspond to 2500-25000 base pairs.) We have 
argued before that $d$ should not be too strictly taken as the dimension of the space 
in which the chain moves. It rather is related to the number of nearest neighbors and the
threshold for enlarging the tube by a new pore. 
\item For longer chains this dependence on $d$ becomes less important and all the 
curves for various $d$ zoom towards a single curve for large $N$. In this sense $d$ 
only influences a correction to scaling, just as the effective exponent can be seen as a
correction to the asymptotic scaling. 
\end{itemize}

It turns out that (apart from the exactly soluble models of the previous sections, where
DMRG gives values indistinguishable from the exact results) DMRG
performs optimally for the pure RD model. Disturbing influences, such as hernias and
barrier crossings, make the DMRG calculation more involved and less accurate. 
We come to this point in the Section \ref{cross}. The behavior in more complicated
situations is still rather similar to that of the RD model.

\subsection{The Diffusion Coefficient}\label{difco}
\vspace*{1cm} 

\begin{figure}[h]
\begin{center}
    \epsfxsize=12cm
    \epsffile{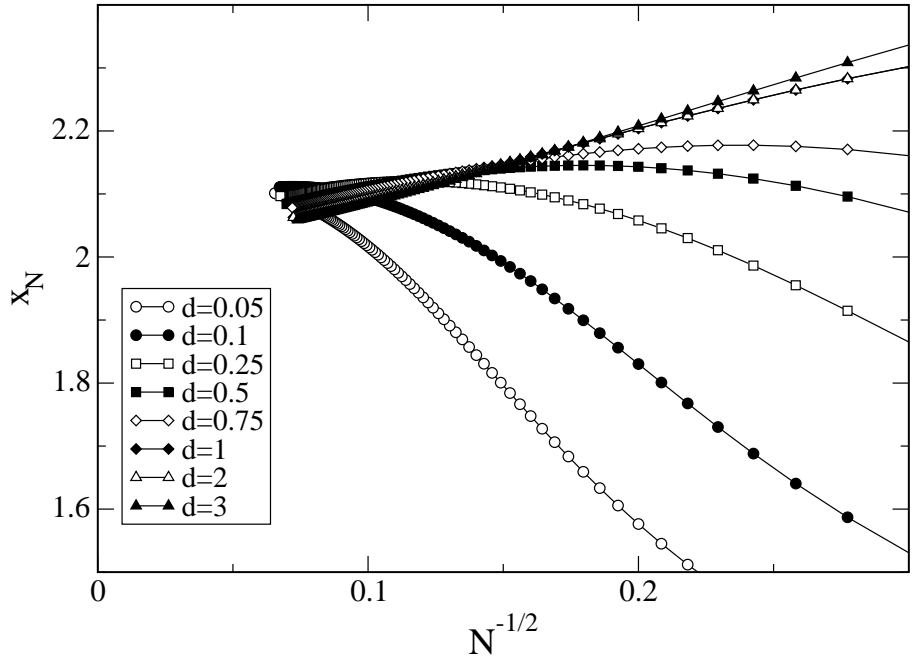}
    \caption{The effective exponent $x_N$ as function of $N^{-1/2}$ for various 
embedding  dimensions $d$.}
\label{diffexp}
\end{center}
\end{figure}

A direct method to obtain the diffusion coefficient is to numerically differentiate the
drift velocity with respect to the field. This can be done for arbitrary fields and yields
interesting behavior of the diffusion as function of the field strength. DMRG data
are accurate enough to do this numerical differentiation, but for stronger fields
the convergence of the method decreases rapidly. The reason is that one has to
find the groundstate of a non-hermitian matrix, for which the Arnoldi method works
best \cite{Carlon1}, as long as the non-hermiticity is not too large.

Again the proper way to determine the power, by which the diffusion decays, is to 
form effective exponents from the values $D(N)$
\begin{equation} \label{h5} 
x_N = {\log D(N+1) - \log D(N-1) \over \log (N+1) - \log N}
\end{equation} 
and to plot them against $N^{-1/2}$. The picture, shown in Fig. \ref{diffexp} is similar
to that for the renewal time: large corrections to scaling and a substantial dependence
on the embedding dimension $d$, with an asymptotic value consistent with the 
theoretical predictions $D \sim N^{-2}$. 

The shape of the curves depends on the power of $N$ which is chosen for the
horizontal axis. We have used here the power $N^{-1/2}$, partly for practical reasons
as it gives the clearest approach to the asymptotical value and partly based on the
behavior of the correlation functions (see Section \ref{localor}), which clearly demonstrate
a region of order $N^{1/2}$ near the ends of the chain, distinctly different from
the bulk.

\section{Cross-over from Reptation to Rouse}\label{cross}

The models discussed so far are all reptative. Characteristic for reptation 
is the large renewal time $\tau \sim N^3$ and the slow diffusion $D \sim N^{-2}$. 
As we saw in the previous section, finite though long chains even have effectively 
larger powers than the asymptotic ones. The common reason for this behavior 
is the persistence of tube configurations. The hopping 
rules only permit the tube to be refreshed from the ends of the chain. Including 
other types of hopping such as hernia creation/annihilation and barrier crossing
may change the asymptotic behavior drastically. These moves, which we collectively
denote with constraint release (CR), can change the tube
internally, sometimes resulting in Rouse dynamics with typical renewal time 
$\tau \sim N^2$ and diffusion $D \sim N^{-1}$. 

Let us assume that the CR moves have a strength $c$. When $c$ is of order unity
the motion will be dominated by this new option in motion and Rouse dynamics will
result. So the interesting case is small $c$, since for $c=0$ the system displays 
reptation behavior. The situation can be represented by the formula \cite{Drzewinski4}
\begin{equation} \label{i1}
\tau (N,c)  = N^3 g (c^\theta N),
\end{equation} 
where $g(x)$ is a cross-over function and $\theta$ the cross-over exponent. $g(x)$ is
supposedly a regular function both for small and large argument. Around $x=0$ it 
has the power series expansion
\begin{equation} \label{i2}
g(x) = g_0 + g_1 x + \cdots
\end{equation} 
with a non-vanishing $g_0$, leading for  $c=0$ to reptation dependence on $N$.
For large argument we expect the expansion
\begin{equation} \label{i3}
g(x) = g_{-1} x^{-1} + g_{-2} x^{-2} + \cdots
\end{equation} 
with a non-zero coefficient $g_{-1}$ in order to yield the Rouse dynamics in this limit.

The cross-over exponent $\theta$ dictates for what combinations of $c$ and $N$ one
can expect either of the two types of motion. We will argue that $\theta = 1/2$
in many cases. Then the turning point occurs for $N \sim 1 / \sqrt{c} $.
Below that  value the behavior is more reptative and above that value it is more 
Rouse-like. Unfortunately the cross-over is not always adequately described in a 
single parameter $c$, notably not in the RD model.

For the diffusion a similar cross-over representation exists
\begin{equation} \label{i4}
D(N,c) = N^{-2} f(c^\theta N)
\end{equation} 
with expansions as in (\ref{i2}) and (\ref{i3}) for the cross-over function $f(x)$.
We expect the cross-over exponent to be equal to exponent of the renewal time, 
because renewal dictates the time scale determining the type of behavior. 

There is a method to directly determine the exponent $\theta$ which is based 
on the large $x$ behavior of the crossover function $g(x)$.
Inserting the asymptotic behavior (\ref{i3}) into (\ref{i1}) we obtain
\begin{equation} \label{i5}
\ln(\tau/N^2) = \ln g_{-1} - \theta \, \ln c + \cdots,
\end{equation} 
where the dots refer to corrections of order $1/N$. First make an extrapolation to
$N \rightarrow \infty$ of the left hand side of (\ref{i5}), which usually can be 
made since the data approach the limit in a fairly linear way. 
Then plot these extrapolated values as a function of $\ln c$. 
If the curve is straight, the slope gives the value $\theta$. When curve is not 
straight,  the slope gives the ``local'' value of $\theta$ as function of $c$. 

As the cross-over behavior in the cage model is simpler, we start the discussion 
with this case.

\subsection{Cross--over in the Cage Model}

In the cage model the basic move is the hernia migration (see Fig. \ref{repchd}). 
Without other hopping mechanisms the model displays reptation. The $d=1$ 
dimensional case is exactly soluble and has been discussed in the section 
Section \ref{hernia1}. Here we discuss the 2-dimensional case, since, 
as yet, $d=3$ turned out to be too hard for obtaining accurate data.

The perturbation from reptation consists of allowing barrier crossings, as
form of constraint release. 
In the original definition of the model these moves have explicitly been 
considered as possible, but their influence were not investigated \cite{Doi}.
As argued above the most interesting scenario results when we turn them on 
gradually with an amplitude $c$. The argument that the hernia migrations
alone lead to reptation is based on the concept of the backbone of the chain. 
We may define the backbone by eliminating systematically all
the hernias in the chain. As the hernia migrations do not change the backbone,
reptative behavior follows because the backbone is a slow variable 
only changing through end-repton motion.
Barrier crossings change the chain internally and in combination 
with hernia migration the tube structure is locally modified. 
Thus any $c$ will induce cross-over to Rouse dynamics. 

DMRG calculations for the cage model are carried out in \cite{Drzewinski5}. 
\begin{figure}[h]
\begin{center}
    \epsfxsize=12cm
    \epsffile{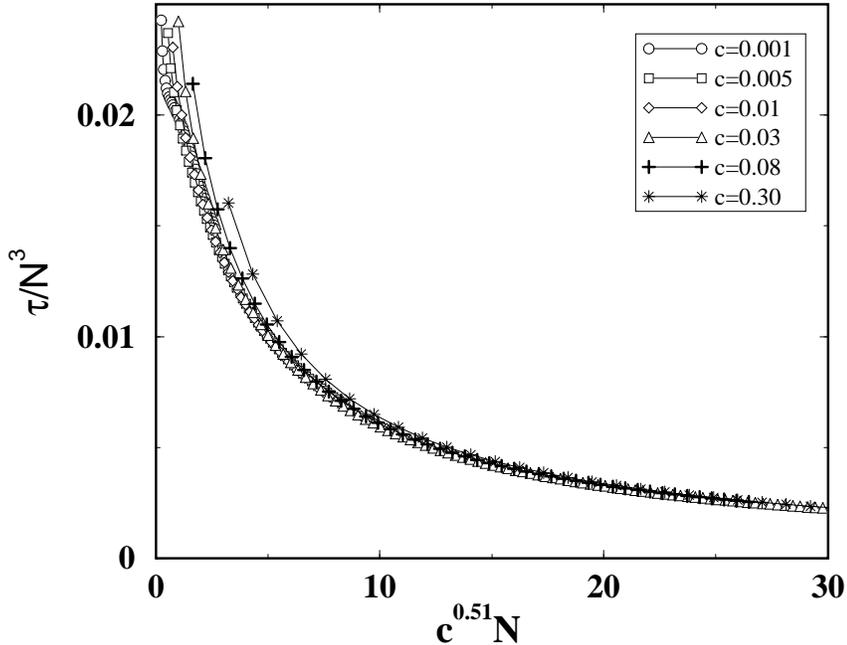}
    \caption{The cross-over function $g(x)$}
\label{crosscager}
\end{center}
\end{figure}
In Fig. \ref{crosscager} the various curves for the renewal cross-over function $g(x)$ 
are plotted. The collapse of the data on a single curve shows that the representation
(\ref{i1}) is fairly adequate to accommodate all data which are collected for $N$ and $c$ 
values. The collapse is based on the choice $\theta=1/2$. The same value of $\theta$
is obtained from the calculation of the local value, which indeed turns out to give a 
value independent of the regime in $c$ where it is determined.

\begin{figure}[h]
\begin{center}
    \epsfxsize=12cm
    \epsffile{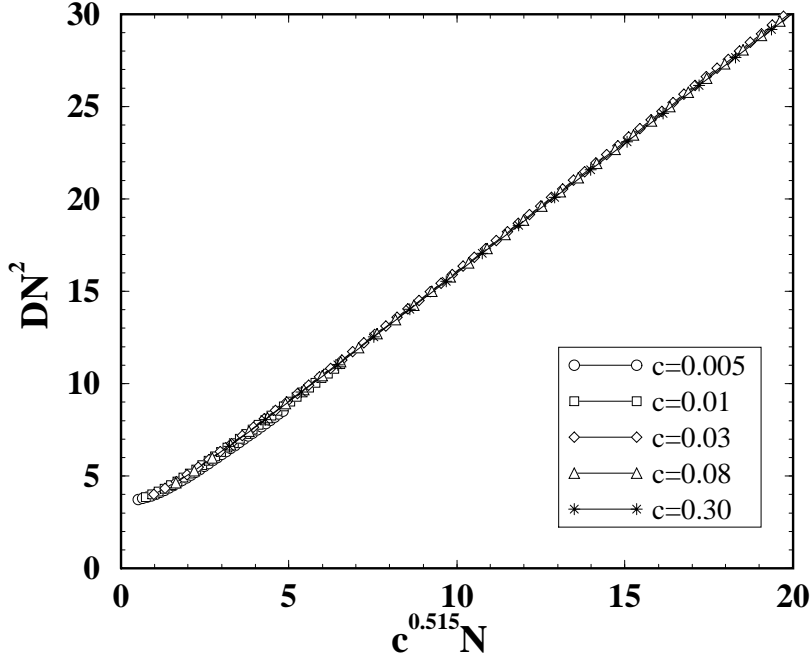}
    \caption{The cross-over function $f(x)$}
\label{crosscaged}
\end{center}
\end{figure}

The diffusion data give an even nicer collapse of the data as Fig.~\ref{crosscaged}
shows.
This proves that a cross-over representation is very efficient to organize the data
for the renewal time and the diffusion, which exhibit a wide variety in the finite 
size regime. 

\subsection{Cross--over in the RD model}

In the RD model, both hernia creation/annihilation and the barrier crossings change 
the internal structure of the tube, but neither of the two will change the dynamics 
drastically when operating alone. Here we briefly review the results as published in 
\cite{Drzewinski6}.

Consider first the hernia creation/annihilation. The backbone is the configuration 
resulting after systematic elimination of all the hernias. When 
hernia creation/annihilation is present the tube will fluctuate with hernias, 
but the backbone of the tube will remain invariant. This backbone can only be 
altered by reptons at the end of the chain. Thus an RD chain with hernia 
creation/annihilation will have again the long renewal time $\tau \sim N^3$ and 
the slow diffusion $D \sim N^{-2}$. The hernia processes speed up the dynamics 
because the backbone is shorter than the tube, but it is still of the order $N$. 
In particular the combination of the hernia creation/annihilation near the ends 
of the chain with the motion of the end reptons removes barriers from the chain, 
but it does not affect the bulk of the chain. The asymptotic behavior of the 
renewal time and the diffusion coefficient will have the 
same exponents as reptation, but it requires larger $N$ to see it.

The barrier crossings do change the backbone of the chain, but without the help of
hernia creation/annihilation, they are not able to change the tube. As one
sees from Fig.~\ref{repchc} a barrier crossing only interchanges two links in 
the sequence of taut links. Moreover two neighboring opposite links cannot 
interchange. Thus the subsequence of links, that are oriented along one axis, 
e.g. either in the positive or the negative $x$ direction, remains invariant 
under barrier crossing. So only in combination with hernia creation/annihilation 
this sequence is changed as two opposing links are created or annihilated.

\begin{figure}[h]
\vspace*{12mm}

\begin{center}
    \epsfxsize=12cm
    \epsffile{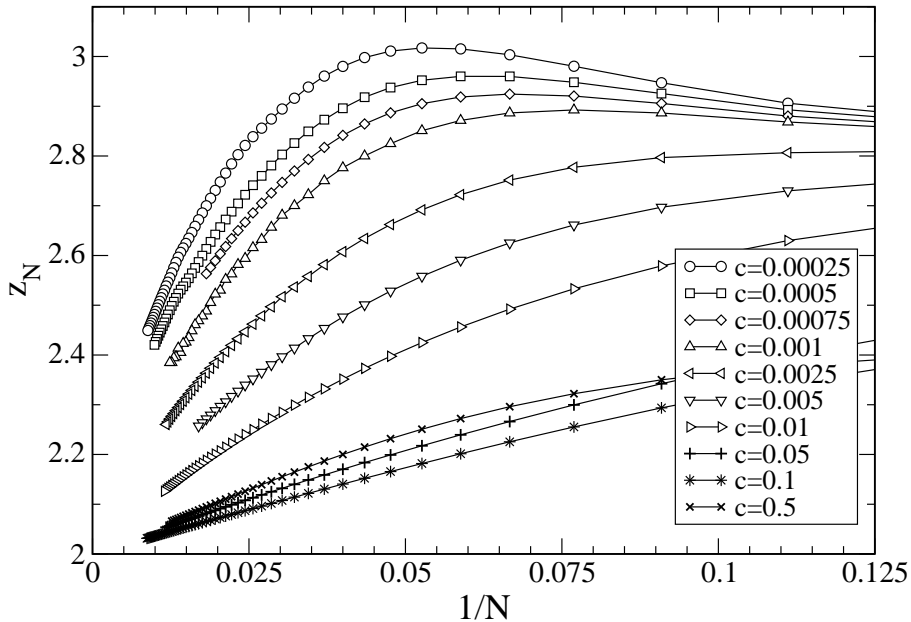}
    \vspace*{5mm}

    \caption{The renewal time exponent $z_N$ for $h=0.1$ and a set of values $c$.}
\label{crossrdc}
\end{center}
\end{figure}

\begin{figure}[h]
\vspace*{5mm}

\begin{center}
    \epsfxsize=12cm
    \epsffile{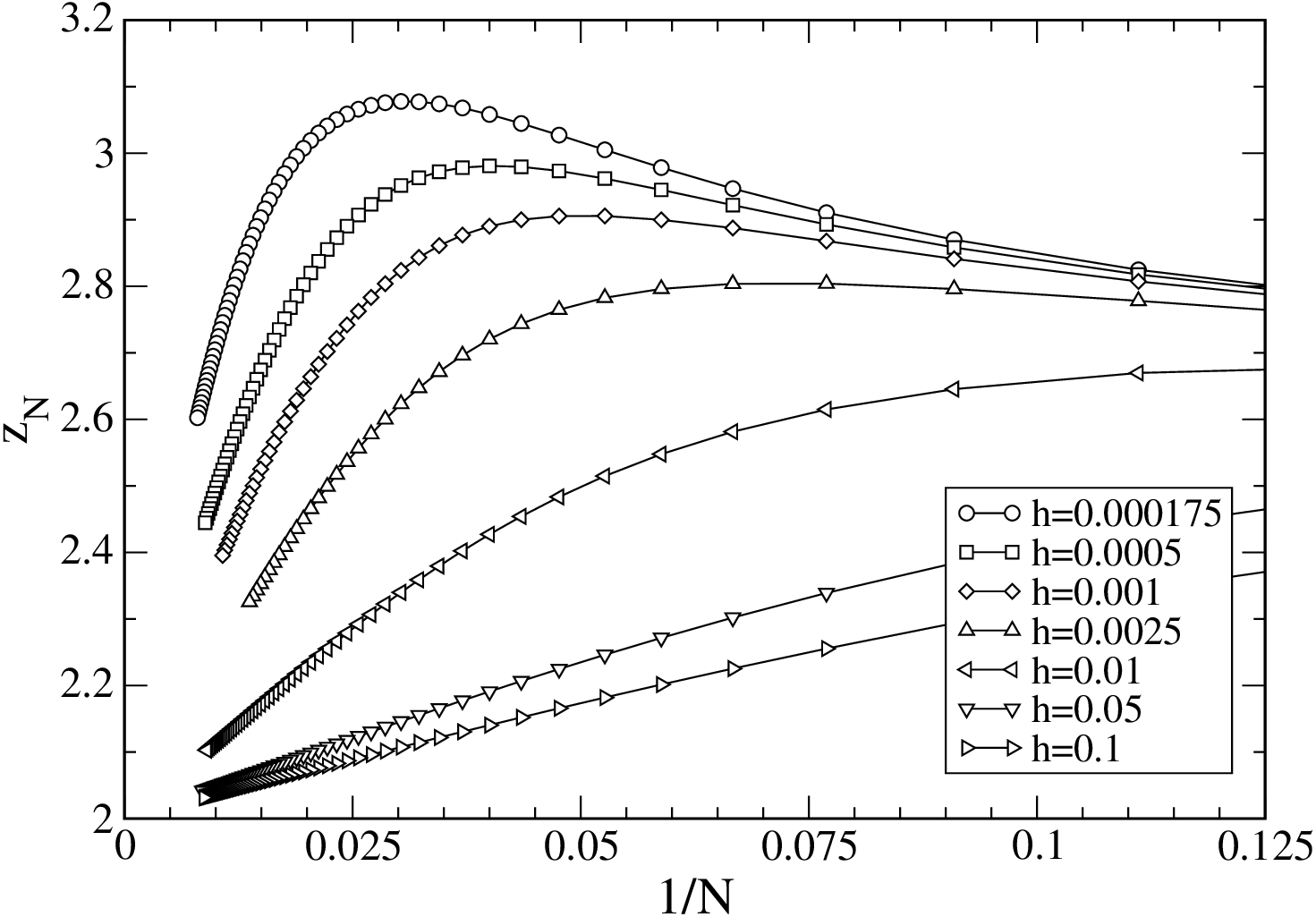}    
     \vspace*{5mm}

    \caption{The renewal time exponent $z_N$ for $c=0.1$ and a set of values $h$.}
\label{crossrdh}
\end{center}
\end{figure}

\begin{figure}[ht]
\begin{center}
    \epsfxsize=12cm
    \epsffile{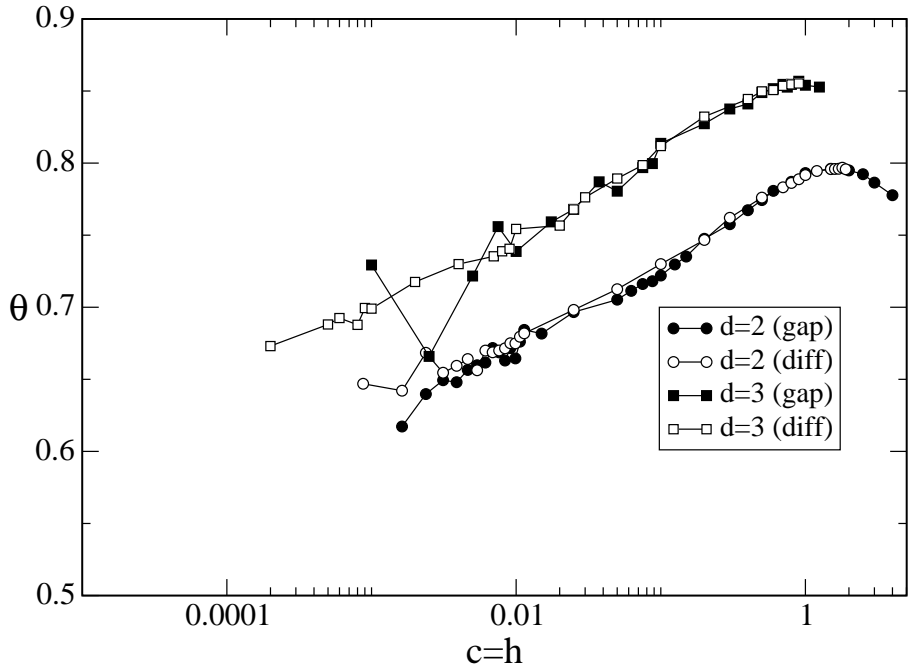}
   \vspace*{5mm}

    \caption{Comparison of the local crossover exponent $\theta$ for $d=2$ and $d=3$, 
both for the gap and the diffusion coefficient.}  \label{compd2d3}
\end{center}
\end{figure}

Thus most interesting cross-over will result when both types of perturbation are 
present. We give barrier crossings the  amplitude $c$  and the hernia
creation/annihilation the amplitude $h$. First we look in Fig.~\ref{crossrdc} 
to the case where sufficient hernia
processes are available ($h=0.1$) and the barrier crossing are slowly turned on. 
One observes the typical behavior of a curve that for small $N$ seems to exhibit 
a reptation exponent $z_N=3$ and for larger $N$ turns over to $z_N=2$. The 
smaller $c$, the more outspoken this trend. 
Next we look in Fig.~\ref{crossrdh} to the opposite case where sufficient 
barrier crossings are available $c=0.1$ and we inspect the behavior for 
small $h$. Now $h$ plays the role of the small parameter in the cross-over 
representation (\ref{i1}). The trend is the same as in the previous picture. 
On the basis of these curves the crossover functions for the renewal time and the 
diffusion coefficient can be determined. A more sensitive test is the determination
of the local value of $\theta$ as a function of the small parameter $c$ or $h$.
The results are somewhat disappointing. Rather than leading to a unique value, 
as in the cage model, one finds a $\theta$ that changes with the value of the 
small parameter. This is most pregnant in the case where the line $h=c$ is 
investigated. In Fig.~\ref{compd2d3} the value of $\theta$ is shown for the line 
$h=c$ both for an two-dimensional and a three-dimensional embedding lattice. 
Also the curves are plotted for diffusion and renewal time. One observes that 
diffusion and renewal involve the same $\theta$. There seems
to be a dimensional effect since $d=3$ leads to a significant larger value of 
$\theta$ than $d=2$. It is interesting that for the RD model with sideways 
motion the case $d=3$ could be calculated with some confidence, 
using diligently the symmetries of the model. 
The real challenge is the behavior for combinations of smaller
crossover parameters $h=c$ and correspondingly longer $N$, because it is there where
the crossover formulae apply. One sees a trend towards the ``universal'' exponent 
$\theta=1/2$, but it is still a long way to go for accurate results.

\section{Local Orientation}\label{localor}

In Section \ref{correl} we indicated how the local orientation of the chain can be deduced
from the probabilities on the configurations of a single link (see (\ref{c4})). The local
orientation tells more about the structure of the chain than the global quantities as the
renewal time and the diffusion coefficinent. Local densities have been  obtained through simulations
by Newman and Barkema \cite{Newman1}. Extensive calculations of local orientations, 
both for the magnetophoresis and electrophoresis versions of the RD model, 
have been carried out using the DMRG method \cite{Carlon1,Drzewinski1,vanLeeuwen3}. 
For weak fields they are interrelated according to relation (\ref{d11}) of Section \ref{linear}.
Typical properties for each of the cases mutually help determining the shape of 
the curve. We start out with the results for weak fields and then discuss the results
for stronger field.

\subsection{Weak Field Orientation} \label{weakloc}

In Fig.~\ref{corfigMP} and \ref{corfigEP} we have shown the curves for the local orientation
of the magnetophoresis and electrophoresis case. It is evident that the electroporesis 
curve is more informative than the magnetophoresis curve. So we comment mainly on the 
former. The most salient feature of the curves in Fig.~\ref{corfigEP}  is the symmetry, 
given by the property (\ref{d8}). It follows of the invariance of the chain under 
transposition i.e.~interchange of head and tail, which is permitted since the 
charges of all reptons are equal. As the orientation
is proportional to $\epsilon$ in the linear approximation, it suffices to give the value 
of the orientation divided by $\epsilon$ for the various lengths of the chain. They are 
given in Fig.~\ref{corfigEP} for $d=1$. 

The next feature to which we want to draw the attention is the long linear stretch in the
middle of the chain for long chains. It points to bulk behavior in contrast to boundary 
behavior at the ends of the chain. Symmetry requires the orientation to be zero in the
middle point of the chain. The average shape of the chain follows as a sum over 
the orientations as function of the position. So counting from the middle the chain 
curves upwards towards the head {\bf and} towards the tail, leading to a U-shaped
chain. Of course this is average behavior and it could be a superposition of
chains which are  only curved upwards near the ends. But inspection of the probabilities
of the configurations for short chains confirms that the U-shaped chain have the
largest probability. So the chain tends to hook around obstacles.

If this tendency of bulk behavior would persist to the ends of the chain we would
get a large orientation for the end segments. However we found 
from (\ref{d32}) in Section \ref{relat} that for small $\epsilon$
\begin{equation} \label{j1}
\langle y_1 \rangle = v_d - {2d \over 2d+1} \epsilon
\end{equation} 
The drift velocity $v_d$ gives for long chains a minor contribution, because it vanishes as
$\epsilon/N$. So the second term determines the value. Therefore all the curves start 
practically as $\langle y_1 \rangle =-2d/(2d+1) =-2/3$ at the tail and 
at the head on the value $2/3$. So the linear behavior of the bulk has to bend
over in order to reach the values prescribed by (\ref{j1}). Physically this is the 
result of the disorder produced by the constant traffic of stored length in and out the 
chain (contour length fluctuations CLF).

The maximum and minimum are a distance of order $\sqrt{N}$ apart from the 
corresponding ends.  This is an important indication for finite size corrections. 
Their leading term may be expected to be of the order $\sim 1/\sqrt{N}$.
The zones of different behavior show
the delicate balance between the disorientation at the ends and the orientation deeper
in the chain (but not too close to the middle where it vanishes because of symmetry).

\subsection{Electrophoresis at larger fields}\label{electro}

The tranposition symmetry (\ref{d6}) of the chain with equal charge is not restricted to 
small driving fields, but holds for any strength of the field. The tendency to form 
U-shaped chains becomes stronger the stronger the field.  Unfortunately
this effect prevents to carry out DMRG calculations for strong fields and long 
chains. The Master Operator becomes very asymmetric (non-Hermitian) and 
the Arnoldi method fails to find the stationary state. Likely this is due to the
fact that stationary state distribution differs very much from the weak field
case where all configurations get practically equal weight. 

This phenomenon is substantiated by the calculations of Kolomeisky and Drzewinski
\cite{Kolomeisky} for small chains, where the individual configurations can be
inspected. They find that initially the drift velocity increases with the driving field,
then flattens off to a maximum and for even stronger fields decays exponentially to
zero. At the same time the probability gets concentrated on U-shaped configurations.
A similar behavior results at constant field for increasing length. 
This effect is a serious restriction of the applicability of gel-electrophoresis to sort
the chain according to their length. The linear increase at weak fields implies that
in a bunch of chains the shorter ones proceed faster in a gel and so the spatial 
distribution in a gel-electrophoresis experiment reflects the length distribution.
The maximum in the drift velocity is known as ``band collapse''; a wide range
of lengths move equally fast in a field. In order to free the chain from their obstacles
the field is often reversed (field-inversion gel-electrophoresis).

\subsection{Magnetophoresis at larger fields}\label{magor}

In contrast to the symmetric chain, the case where the charge is only at the head,
yields a positive orientation everywhere in the chain. 
If one pulls the chain by the head repton only, the head link gets strongly oriented and
this in turns orients the next link etc. The orientation will be 
monotonically decreasing towards the tail, as the pull at link $j$ is mediated by 
the links $k>j$ and due to fluctuations it decreases steadily. 

The first observation is that the expression for the curvilinear velocity  $J$ simplifies 
\cite{Barkema1}. We have derived from  equations (\ref{e15})--(\ref{e18}) that the 
probability on a slack link $p^0_j$ increases linearly with the position $j$ and that the 
slope is given by the curvilinear velocity. We copy the last equation, relating the 
probabilities on the end links with the curvilinear velocity 
\begin{equation} \label{j2}
p^0_N = p^0_1 - (N-1) J.
\end{equation} 
The end-repton contributions to the curvilinear velocity give additional information
on the end-link probabilities. The tail repton yields for $B_0 = 1$ (see (\ref{d29})) 
\begin{equation} \label{j3}
J_0 = 1 - (2d+1) p^0_1 = J.
\end{equation} 
The two relations (\ref{j2}) and (\ref{j3}) are two equations involving the three unknowns:
$p^0_1, p^0_N$ and $J$, one short of solving them. We have also information from the 
general expressions for the drift and curvilinear velocities in terms of the probabilities
of the end links. 
For the head repton we do have a $B \neq 1$ and from (\ref{d29}) we obtain
\begin{equation} \label{j4}
J_N = (d p^0_N - p^-_N) B - (d p^0_N - p^+_N) B^{-1} =J.
\end{equation} 
The drift velocity gives also information on the end-link probabilities. (In the bulk one
cannot eliminate the two-link correlations, as we could in (\ref{e15}).) So we have two 
equations : one for the tail-repton drift-velocity (in magnitude)
\begin{equation} \label{j5}
v_0 = p^+_1 - p^-_1 =v_d
\end{equation} 
and the other one for the head repton
\begin{equation} \label{j6}
v_N = (d p^0_{N} + p^-_N) B - (d p^0_{N} + p^+_N) B^{-1} = v_d.
\end{equation}
In addition we have two normalization equations (\ref{c2}) for the link probabilities
\begin{equation} \label{j7}
p^0_1 + p^+_1 + p^-_1 =1, \quad \quad  {\rm and} \quad \quad p^0_N + p^+_N + p^-_N =1.
\end{equation}
Counting the number of unknowns, we have the three link probabilities for tail and head and 
two velocities $v$ and $J$, while we have seven equations ((\ref{j2}--(\ref{j7})), 
still one short  of solving the system. The situation is similar in the RD model 
with equal charges, where one has $J=0$, but fails to have (\ref{f2}).
\begin{figure}[h]
\begin{center}
    \epsfxsize=12cm
    \epsffile{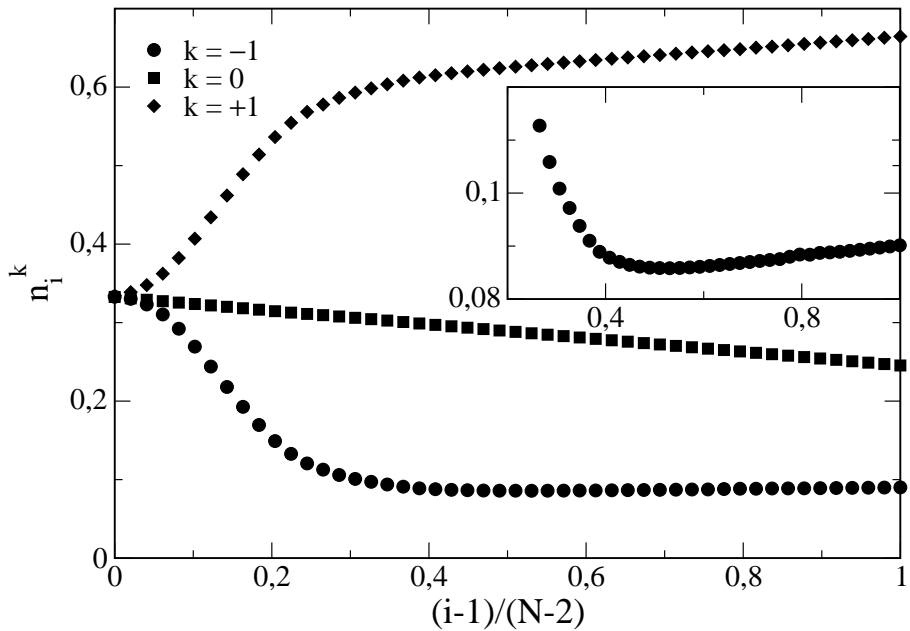}
	\vspace*{2mm}

    \caption{The probabilities for a link $j$ at $\epsilon=1$. The upper curve, representing
$p^+_j$ is monotonously increasing, the middle curve the trivial straight $p^0_j$ and
the lower curve refers to $p^-_j$; it is non-monotonous as the inset shows clearly.}
\label{densE1}
\end{center}
\end{figure}

For stronger fields the chain gets more oriented. For a fully oriented chain 
the curvilinear velocity equals the drift velocity. 
In fact, putting them equal, $v_d = J$, supplies the lacking equation.
It  gives an excellent approximation for drift velocity as function of the field and 
chain length \cite{Drzewinski1}. The ansatz fails for weak fields 
where the drift velocity is linear in $\epsilon$, while the curvilinear velocity is 
quadratic in $\epsilon$ (reversal of the field has no influence on the curvilinear 
velocity.)

As (\ref{j2}) shows, the distribution of slack links is linear for this
model. That does not help to understand the orientation, which is the difference 
between the up and down taut links. Without detailed calculations it is obvious that
the orientation will be strongest at the head, which is pulled by the field. So the last 
link will be taut and up. This tendency propagates towards the tail with decreasing 
amplitude. At the tail, where the influence of the field is weakest, all three possibilities
will be equally present (for $d=1$).
In Fig.~\ref{densE1} the individual probabilities $p^k_j$ are plotted 
for the (strong) field strength $\epsilon =1$. The middle curve is the trivial $p^0_j$, the
upper and lower curves refer to $p^\pm_j$. The interesting feature is that $p^-_j$ is 
non-monotonuously decreasing, while $p^+_j$ increases steadily. In \cite{Drzewinski3}
this unexpexted behavior is explained by considering the production of taut links and 
following their history.

\section{Remaining Problems}\label{outprob}

The physics of polymers is so rich that many problems remain, even in the context 
of lattice dynamics. Although the class of hopping rules considered here is fairly broad,
the lattice models have many shortcomings with respect to the real polymer dynamics. 
In this review we have widened the class of original models by considering combinations
of hoppings that yield a spectrum of dynamical behavior ranging from reptation to Rouse 
dynamics. Of course it would be nice to further extend the possibilities such as to include 
more details of the motion. For instance the models lack the influence of hydrodynamic 
interaction between parts of the polymer, which is relevant for polymers in good solvents.
Such mechanisms are, however,  at odds with the lattice description, because they tend 
towards collective motion of the reptons. So there is no point of trying to include them. 

Another shortcoming, alluded to in the introduction, 
is the lack of self-exclusion between the parts of the chain. We have argued that such an 
effect is less severe on the repton level than on the monomer level. Moreover the
mechanism of self-avoidance is not present in melts. 
So the non-interacting chain is not only a
theoretical concept but it also is of experimental relevance. The problem has two aspects: 
on short-range and on long-range distance. The effect on the short
range is that successive reptons cannot unlimitedly occupy the same cell (as is allowed
by our lattice models). This is likely not of influence on the universal behavior of 
long chains. A modest step in the direction of self-avoiding reptons was taken by Kooiman 
and van Leeuwen \cite{Kooiman1,Kooiman2}. They considered chains where no more than two 
successive reptons are allowed in the same cell. Without finding an expression analogous 
to (\ref{f2}), they deduced for periodic chains for the diffusion coefficient
\begin{equation} \label{k1}
D \simeq {(2d+3) d[1+2/(d(2d+3)^2]^{1/2} -1] \over N^2} \, \simeq {1 \over (2d+3) N^2 }. 
\end{equation} 
The second approximation is only meant for comparison with (\ref{f5}). 
This results has the same status as (\ref{f5}), but so far it has not been backed up 
by a rigorous proof (or falsification). Note
that (\ref{k1}) differs from (\ref{f5}) not in the power of $N$, but in the prefactor
One can consider the exclusion of two
successive reptons in the same cell as a form of stiffness of the chain. This holds also
for the necklace model where backtracking of the chain is forbidden. The notion of
reptons, however, already includes some stiffness of the chain. Note
that (\ref{k1}) differs from (\ref{f5}) not in the power of $N$, but in the numerical
constant. 

The mutual exclusion of distant parts of the chain is the more severe problem. 
It implies an interaction between all parts of the chain. In the
DMRG method it is hard to include such interactions.
The long-ranged interactions due to self-exclusion change the universal properties in a 
non-trivial way. Self-avoiding chains have a radius of 
gyration $R_g \sim N^\nu$ with $\nu=0.58$ for $d=3$, while $\nu=1/2$ for non-interacting 
chains (in all dimensions). It is of major theoretical interest how this (small) deviation 
of $\nu$ affects the behavior of other exponents. For instance the
combination $D \tau$ should scale as $R^2_g$ as indicated in (\ref{0}). 
Thus at least one of the exponents $x$ or $z$ has to change its value due to self-avoidance. 

A problem that can be approached within the context of lattice dynamics is the
behavior for stronger driving fields.
In this review we have mainly focussed on the drift velocity for weak driving fields, 
the regime where the drift is proportional to the field strength. The next regime has
values of $\epsilon$ and $N$, where the combination $\epsilon N$ is of order unity. 
Simulations suggest that $\epsilon N$ is the proper combination for crossover to different
dynamical behavior \cite{Barkema3}. 
We like to connect this issue with the physical phenomenon
that for larger fields (at fixed length $N$) the chain does not reorient itself anymore. 
While for small values of the field the head and tail frequently change role as foremost 
repton in the field direction, such a complete reversal gets very rare for larger fields 
(or for longer chains at fixed field). This is a sort of spontaneous symmetry breaking 
with respect to orientation. The effect was specifically investigated by Aalberts and 
van Leeuwen \cite{Aalberts1} in a variant of the RD model called the Fast Extron Limit,
on which we comment in more detail below since it raises an interesting question. 
In the other subsection we speculate how the crossover findings can be used to discuss
the similar problem in polymer melts.  

\subsection{Fast Extron Limit}\label{fel}

This is the limit where the hopping of the external reptons becomes very slow with respect
to the internal reptons. Physically this can be realized by heavy endgroups. 
As in the case of a periodic chain, it is useful to employ the description 
in terms of the slack link or extron occupation $n_l$ (see the discussion in Section
\ref{exact}) and tube configuration ${\bf S}$.
The tube changes on a slower time scale than the occupation and
therefore the extrons come to equilibrium with the momentary configuration of the tube.
To find the equilibrium distribution of the extrons we define a tube potential
\begin{equation} \label{k2}
V_j ({\bf S}) = \epsilon \sum^L_{i=j+1} s_i,
\end{equation} 
which is the potential at cell $j$ in the tube as measured with respect to the cell of the 
head repton. The extrons obey Bose statistics, because $n_l$ extrons in a cell count for 
one configuration, since their order is fixed. (There would be $n_l !$ configurations if 
they could be permuted in a cell.) The Bose-Einstein distribution is most easily given with 
the aid of a thermodynamic potential $\mu({\bf S})$, yielding the average over extron 
occupations as
\begin{equation} \label{k3}
\overline{n_l} ({\bf S})= \left[ {1 \over \exp(V_l ({\bf S})- \mu({\bf S})) - 1} \right].
\end{equation} 
$\mu({\bf S})$ is fixed by the requirement (\ref{f1}) that the total 
number of extrons is given by $N-L$. Strictly speaking one has to expand the 
expression (\ref{k3}) in powers of $z = \exp \mu$ and keep the power $N-L$. 
Because $N-L$ is a large number, it is a good
approximation to determine $\mu$ such, that the average total occupation equals $N-L$.

The dynamics of the chain is now restricted to adding or subtracting a taut link to
the tube. If cell $0$ is occupied ($n_0>0$) we can add a link to the tail and when it is 
empty ($n_0=0$) the tail link can be removed. The same holds for the head cell $L$.
As (\ref{k3}) implies the occupation probabilities, one knows the transition rates for
the Master Equation for the tube configurations ${\bf S}$. This Master Equation has
been simulated \cite{Aalberts1} and it allows to handle fairly long chains efficiently, 
notably by an accurate approximation to determine the thermodynamic potential, 
which is the most time consuming step in the procedure. Note that, in contrast to  
periodic chains, the tube length $L$ is not a fixed parameter but fluctuates in every 
step (either shortening or growing the tube). For zero field the distribution $P(L)$ is
Gaussian around the value $L=2dN/(2d+1)$ and it keeps this shape for reasonable 
field strengths. 

Quite different behavior follows for the distribution $P(S)$, where $S= \sum_l s_l$ is
the tube/chain orientation in the field direction. For small 
fields it is again Gaussian, but for stronger fields it develops two peaks. This is 
strongly reminiscent of the behavior of the magnetization in a magnetic field. For high 
temperatures it is Gaussian distribution, but below a critical value it develops two sharp
peaks around plus or minus the value of the spontaneous magnetization. However there are 
differences with respect to the thermodynamic limit. Whereas in magnetic systems one 
can take the thermodynamic limit and get a sharp transition point between the two
phases, the thermodynamic limit cannot be reached in a simulation of polymer chains.
In a magnetic system the valley between the peaks becomes exponentially small. For
a reptating chain this was not observed. It was found that the double peak existed for
\begin{equation} \label{k4}
\epsilon N^{1.19 \pm 0.01} > 8.3,
\end{equation} 
as the criterion was taken to be the appearance of a valley between the peaks. 
The relation has been checked for chain lengths $50 < N < 3000$. The emergence of a 
power 1.19 is somewhat surprising, since usually one gets multiples of 1/2 as 
characteristic power behavior. It differs also from the estimate of Widom and Barkema
\cite{Widom,Barkema3}, who give 1 for the exponent. The limit $\epsilon \rightarrow 0$ 
and $N \rightarrow \infty$ would lead to a sharp transition. Finally it is noteworthy that 
fast extron limit permits a systematic expansion in 
$\epsilon$, which can be solved term by term \cite{Aalberts2}.

The upshot of this simulation is that similar behavior is to be expected in the RD model. 
The physical mechanism of the ``orientation transition'' is quite clear. For weak field
and short chains, the chain frequently changes orientation, with sometimes the head
as foremost repton in the field direction and sometimes the tail. When the chain gets
longer, these flips in orientation take more time and become rare.
It can be observed by monitoring the value of $S$ during the simulation.
One may wonder why this problem has not been investigated in detail in the RD model,
using the DMRG method.  
The reason is that the matrix of the Master Operator becomes rather asymmetric, 
which makes the determination of the stationary state slow and unreliable for longer
chains. But it certainly would be desirable to know the behavior for the limit 
discussed in (\ref{b11}). 

\subsection{Polymer Melts}

Lattice models are primarily suited for reptation in a gel. As long as the gel is
assumed to be a rigid network, there is no room for contraint release. Since gels
are not perfectly rigid, one can see a role for the sideways motion that we have
discussed in this review. Constraint release is the central issue in polymer melts.
Usually the same length polymers form the environment for the test polymer. 
Then the amount of constraint release depends on the length of the polymer.
In our lattice models we take the strength of 
sideways hopping rates independent of the length $N$. So these results cannot
be directly applied to the melt.
However, it is conceivable and in practice carried out
by Smith et al. \cite{Smith2} and by Zamponi et al. \cite{Zamponi}, 
to study a tracer polymer of length $N$ in a
melt of polymers of length $N_m$. Then one can fix the contraint release 
rates on the basis of given value of $N_m$ and then vary $N$. To this approach
the calculations given in this review apply. It requires a well founded 
dependence of the constraint release rates on $N_m$.
Once this program has been completed, one can apply it to the equal length
melts by selecting the cases $N=N_m$. Needless to say that this will be a rather
time consuming enterprise. Also it presupposes that the influence on the 
surrounding polymer can be fully represented by a constraint release rate such that
this ``mean field'' approximation is adequate, which has been doubted recently
\cite{Panja}.

{\bf Acknowledgements}. A review like this has benefitted from discussions with
various people having been involved in all stages of the work. 
Of the many people who have been
instrumental in shaping this review, we want to mention B. Widom who has, by 
his lucid lectures, interested one of us (J.M.J.van L.) in the subject many years 
ago. Enrico Carlon was the pioneer in getting the DMRG calculations started
and the main force behind the interpretation of the  results. Gerard Barkema and
Debrabata Panja were
inspiring listeners in the later stages of the work and, by their wide knowledge 
of polymer physics, helped to put the theory into a proper perspective. 
A.D. thanks the Wroclaw Centre for Networking and Computing for access
for their computing facilities (grant No. 82).
\appendix

\section{1-dimensional reptation with hernias} \label{1dhernias}

With the technique of contracting the Master Equation, described in Section
\ref{contra} we can articulate the map of the herniating chain of section \ref{hernia1}.
The empty link and the doubly occupied link are 
identified as a slack link $y_i=0$ and a $+$ particle as a taut link
$y_i=1$ and a $-$ particle as a taut link $y_i= -1$. The hopping of particles corresponds
to the interchange of taut and slack links, as in the RD model.
Interesting is the case where an empty link neighbors a doubly occupied link. 
Both are slack in this map. But the doubly occupied
link may donate a particle, either $+$ or $-$, to the empty link, thereby creating two
opposite taut links or a hernia. The reverse process is the annihilation of a hernia.
The bias in the moves of the particle systems is taken the same as the RD model.

The contraction goes from a $4^N$ dimensional configuration space of the combined 
$+$ and $-$ system to the $3^N$ dimensional configuration space of links.
As we have pointed out in Section \ref{contra} one needs a relation between the
probabilities of the larger space in order to make the contraction useful.
We make an ansatz and later investigate its validity. As mentioned, the driven
systems of $+$ and $-$ particles are the same if also the driving field for the $+$ system 
is opposite to that for the $-$ system. This follows from  a transformation in which
the holes of the $+$ system are mapped into the particles of the $-$ system and vice versa.
In this map an empty link transforms into a doubly occupied link.
Thus we make the ansatz that the probability on an empty link
is the same as that on a doubly occupied link in the $+$ and $-$ system. Then we
can combine the weights and deduce the effective transition rates for the contracted
system as indicated in Section \ref{contra}.

The new rules are a bit complicated and we list the various cases.
\begin{itemize}
\item Transitions of a slack-taut combination. The slack link corresponds to 2 states
of the $+$ and $-$ particle system. Each of them can interchange the taut and slack link,
so the transition rate is the same.
\item Transitions of a slack-slack combination. This corresponds to 4 states of the 
particles. Two of them cannot move: the empty-empty and the double-double 
combination. The two others can create a hernia. So the hernia creation rate is 1/2.
\item Hernia annihilation. A hernia corresponds to an unique particle state and it may
develop into a slack-slack pair in two ways: the $+$ particle may move or the $-$ 
particle. So the hernia annihilation rate is 2.
\item A slack end-link. It has 2 particle configurations, empty and doubly occupied, 
and both may move to the taut position, 
either by creating a particle in the empty link or by annihilation of a particle
in the doubly occupied case. So the transition rate is 1.
\item A taut end-link. It is a unique particle state and it can transform itself in two
ways in a slack state: by annihilation of the particle or by creation a particle of the
other kind. So it has a transition rate 2.
\end{itemize}
So  the particle system maps onto a chain with asymmetric transition rates for the
end reptons and for creation and annihilation of hernias. The other moves are the same
as in the RD model. It may be a bit surprising that, even without a bias, the transition
matrix is non-symmetric. As the map on the particle system shows, it must be a removable
asymmetry and it does not introduce a new lack of detailed balance (as is due to the bias 
in the system). It is similar to the case of the d-dimensional RD model, which has an 
asymmetric transition matrix, but has an an embedding in a larger symmetric system.

In fact the matrix can be symmetrized in exactly the same way as the RD model can 
be symmetrized (for $B=1$). Consider the function
\begin{equation} \label{f9}
\Phi ({\bf Y}) = \left( \sqrt{2} \right)^{N-L}
\end{equation} 
i.e. each configuration is given a power of $\sqrt{2}$ equal to the number of slack links
($L$ is the number of taut links). The similarity transformation 
\begin{equation} \label{f10}
{\cal M}' ({\bf Y} | {\bf Y}') = (\Phi ({\bf Y}))^{-1} {\cal M} ({\bf Y} | {\bf Y}') \Phi ({\bf Y}') 
\end{equation}
symmetrizes the matrix ${\cal M} ({\bf Y} | {\bf Y}')$ of the original problem. The
transition rate of a hernia annihilation was 2 and it  now becomes 1, since the ratio of
the $\Phi$'s in the initial and final configuration is $1/2$. Similarly the rate of  reverse
process, the hernia creation, changes form $1/2$ to 1. For the end reptons the
transition from a slack link to a taut link had a rate 2, which is corrected to $\sqrt{2}$,
because one slack link is lost. In the other direction the transition rate is boosted from
1 to $\sqrt{2}$. The moral is that one has a model with hernia creation and annihilation
rates 1 (the same as RD moves) and with an increased rate for the motion of the 
end reptons. Here again one sees that the ratio of slack to taut links is controlled by the
rate of change of the end reptons.

The model has been analyzed by systematic expansion in powers of the driving field 
\cite{Sartoni}, before it turned out to be soluble.
\begin{figure}[h]
\begin{center}
    \epsfxsize=12cm 
    \epsffile{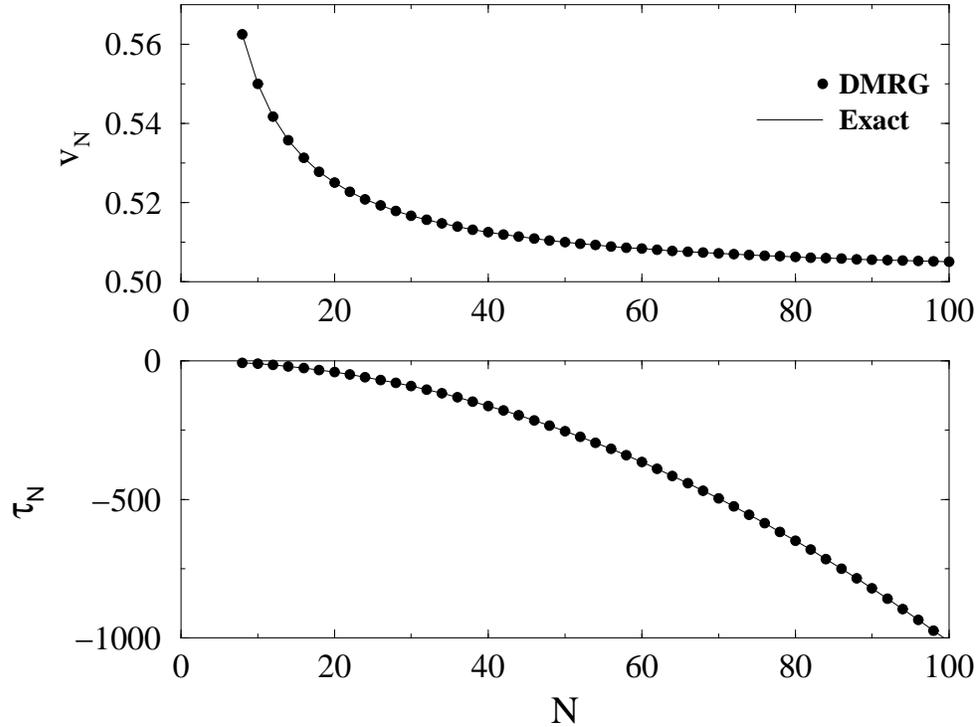}
    \caption{The drift velocity for the RD model with hernias as a function of the length of 
the chain, together with the expression (\ref{f6}) for the necklace model (a). Part (b) 
gives the renewal time of the RD chain with the expression (\ref{f7}) deduced from the 
necklace model}  
\label{compar}
\end{center}
\end{figure} 
Now one still has to verify how good the ansatz is, which enabled the contraction of the
Master Equation. It was based on a global symmetry 
and we used it for the local identification of the probabilities for an empty and a doubly
occupied link. For short chains it does not hold in 
general, but it does to linear order in the driving field. Thus it holds for the gap, 
which is a field free quantity and also for the diffusion coefficient, which stays within
the linear order in the field. The exact expressions for the particle systems are for 
the diffusion coefficient
\begin{equation} \label{f11}
D \simeq {1 \over 4 N},  
\end{equation} 
and for the gap $\Delta$
\begin{equation} \label{f12}
\Delta  = - 2 (1 - \cos (\pi/N)), 
\end{equation} 
which goes asymptotically as $N^{-2}$, leading to a renewal time $\sim N^2$.

As the particle system is ideally suited for the DMRG calculations, we give in 
Fig. \ref{compar} the somparison of the outcome of this calculation for the 1-dimensional
system with hernia creation and annihilation and the above expressions. 
The figure shows that the ansatz is fulfilled and that the DMRG calculations are extremely
accurate.

\end{document}